\documentclass[conference,compsoc]{IEEEtran}

\ifCLASSOPTIONcompsoc
  \usepackage[nocompress]{cite}
\else
  \usepackage{cite}
\fi

\usepackage{tikz}
\usepackage{amsmath}

\usepackage{xcolor}
\usepackage{amsfonts}
\PassOptionsToPackage{hyphens}{url}\usepackage{hyperref}
\usepackage{color}
\usepackage{graphics}
\usepackage{graphicx}
\usepackage{url}
\usepackage{listings}
\usepackage{multicol}
\usepackage{multirow}
\usepackage[scaled]{helvet}
\usepackage{rotating}
\usepackage{xspace}
\usepackage[ruled,vlined]{algorithm2e}
\usepackage{comment}
\usepackage{enumitem}
\usepackage{amsmath}
\usepackage{mathrsfs}
\usepackage{cancel}
\usepackage[capitalise]{cleveref}
\usepackage{subcaption}
\usepackage{fancyhdr} 
\newcommand{\name}{DOLOS\xspace}

\begin{document}

\author{{Aditya Arun, Vaibhav Anand, Wei Sun, Roshan Ayyalasomayajula, and Dinesh Bharadia}\\
UC San Diego}

\title{\Large \bf DOLOS: Tricking the Wi-Fi APs with Incorrect User Locations}

\maketitle

% !TEX root = paper.tex
\begin{abstract}
Wi-Fi-based indoor localization has been extensively studied for context-aware services. As a result, the accurate Wi-Fi-based indoor localization introduces a great location privacy threat. However, the existing solutions for location privacy protection are hard to implement on current devices. They require extra hardware deployment in the environment or hardware modifications at the transmitter or receiver side. 
To this end, we propose \name, a system that can protect the location privacy of the Wi-Fi user with a novel signal obfuscation approach. \name is a software-only solution that can be deployed on existing protocol-compliant Wi-Fi user devices. We provide this obfuscation by invalidating a simple assumption made by most localization systems -- "direct path signal arrives earlier than all the reflections to distinguish this direct path prior to estimating the location". However, \name creates a novel software fix that allows the user to transmit the signal wherein this direct path arrives later, creating ambiguity in the location estimates. 
Our experimental results demonstrate \name can degrade the localization accuracy of state-of-art systems by $6 \times$ for a single AP and $2.5 \times$ for multiple AP scenarios, thereby protecting the Wi-Fi user's location privacy without compromising the Wi-Fi communication performance.

\end{abstract}

% \keywords{Privacy, RF sensing, Localization, Wireless channel, Obfuscation}

\section{Introduction}

% Para 1 -- Wi-Fi based localization has been recongozed and imprtatn for decimeter accurate lcoation.
% The research started at a few meters accurate to now being decimeter acccurate.
User's location in indoor scenarios is an important aspect of providing a personalized experience.
It provides context to a user's activity~\cite{ayyalasomayajula2020deep}, aids in navigation and wayfinding~\cite{ayyalasomayajula2020locap}, and paves the way for smart building automation~\cite{arun2022p2slam}.
For these applications and many more, localization accuracy on the order of tens of centimeters is key.
This led to accurate Wi-Fi-based indoor localization research for more than two decades that started with signal strength (RSSI) localization systems that can provide only a meter-accurate localization~\cite{radar,zee} to many recent systems~\cite{spotfi, ayyalasomayajula2020deep, wipeep, vasisht2016decimeter,luo2023meta2locate} that provide tens of centimeters-accurate user location using Wi-Fi signal's physical-layer information. 

% Para 2 -- Rssi based meter level accurate systems have been improved by the current works achieve decimeter accurate by doing aoa, tof and do not require consent, hence privacy concerns.
Wi-Fi-based localization algorithms primarily rely on three measurements to determine a user's location in these scenarios - the received signal strength (RSSI), the time of flight (ToF) of the signal, and the angle of arrival (AoA) of the signal. 
Concerningly, many of these systems can operate without the consent of the user due to two main reasons: (i) These measurements for a user device transmission happen at the WI-Fi AP, and (ii) A mixture of ToF (802.11mc, 802.11az~\cite{802.11az}) and AoA measurements can be successfully used to localize a user to decimeter-level accuracy and invade their location privacy.

On the other hand, RSSI-based systems commonly provide accuracy in the order of meters~\cite{radar, ez}, meaning users can be localized in different rooms or even different floors. This level of accuracy does not pose a threat to a user's privacy. 
So, in this paper, we aim to deteriorate current AoA and ToF-based localization systems' accuracy form decimeter-accurate to meter-accurate (or as bad as RSSI-based localization systems). Further, a perfect defense should not need any additional hardware or firmware changes, and also not disrupt the user's internet connectivity through the Wi-Fi network.

\begin{figure}[t]
    \centering
    \includegraphics[width=0.85\linewidth]{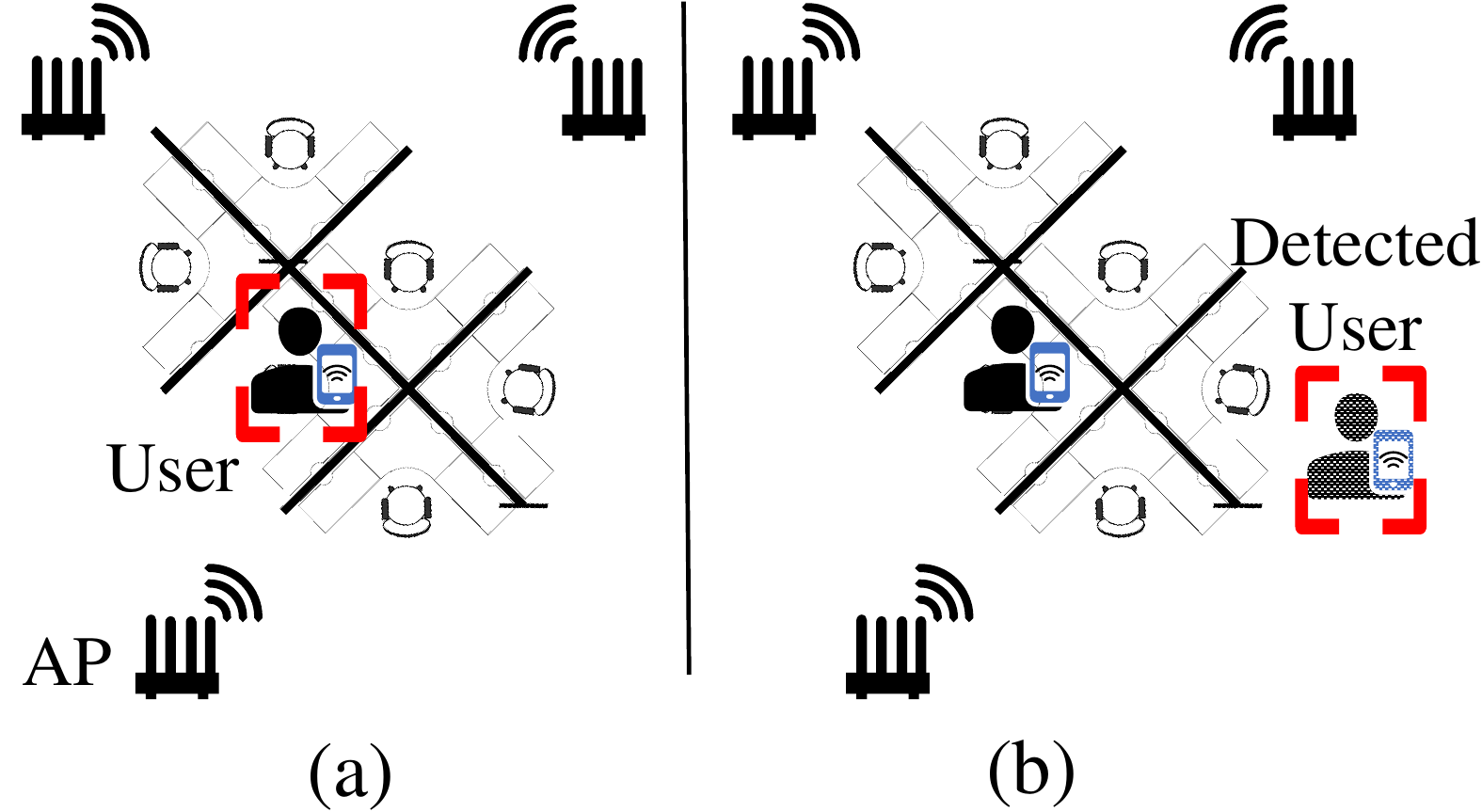}
    \caption{(a) \name-disabled Wi-Fi user can be accurately localized by the Wi-Fi APs thereby introducing the location privacy threat. (b) \name-enabled Wi-Fi user can obfuscate the wireless signals for location privacy protection.}
    \label{fig:main-intro-fig}
\end{figure}

% Para 3 -- Some works for privacy falls short of doing a good job.
Current ToF-based localization systems try to ensure that user can alone make their measurements, and the latest Fine-Grained Positioning (FGP, 802.11az)~\cite{802.11az} proposes a more secure FTM measurement; however, the attack model in WiPeep~\cite{wipeep} demonstrates an attack on 802.11mc FTM, that would still translate to the current state-of-the-art wireless sensing systems.
On the other hand, a few works have explored disabling an AP's ability to localize altogether, by deploying a repeater~\cite{yao2018aegis} or Smart surface~\cite{staat2022irshield} in the environment, tuning the signal's phase and amplitude to prevent a receiver from localizing the user. 
Although these systems prevent the Angle of Arrival (AoA) determination, they require additional hardware to be deployed and do not provide a scalable solution for all spaces. 

% Para 4 --We do a good job by making the decimeter systems as inaccurate as rssi-systems or accurate to only a few meters. Our system is also protocol-compliant and communication preserving. So let us understand what makes this challenging.
We design \name, a system to protect users from being localized without consent in the public or enterprise networks. \name's defense is against the attacker that either could compromise one or more access points (APs), within the public or enterprise networks.
\name's defense model is to deteriorate the AoA and ToF-based localization accuracy of the Wi-Fi localization algorithms to be as inaccurate as RSSI-based localization, i.e., an average of 2-3m median accurate location estimation.
Additionally, \name's design uses existing precoding protocols in Wi-Fi~\cite{beam_forming_doc} to precode the signal prior to transmission at a software level to obfuscate the AoA measurement and build a broadly applicable solution. 
This allows \name to readily integrate into most existing Wi-Fi devices and provide user-location privacy with only a software fix.
Finally, to ensure user's location privacy without compromising their connectivity to the Internet, we designed \name to have no disruption to the existing communication link.

% Para 5 -- First challenge is that the APs have multiple antennas that can resolve multiple paths through aoa estimation, and identify the direct path using the notion that it is the least travelled and most consistent path.
The first design challenge \name attempts to resolve is that, while carrier frequency-based phase-shifts can achieve decent ToF unreliability~\cite{olafsdottir2017security,riaz2022security,staat2022analog}, APs designed for MIMO communication have multiple antennas that can overcome them by performing AoA measurements. As shown in Figure~\ref{fig:main-intro-fig}(a), state-of-the-art Single-AP based decimeter-accurate localization algorithms~\cite{vasisht2016decimeter, soltanaghaei2018multipath,liu2021rfloc,shoudha2023wifi,ma2019wifi,802.11az} rely on a simple fact when they use channel state information (CSI) to estimate accurate ToF and AoA -- "direct-path travels the least amount of time in the air". This helps the algorithms to distinguish the direct path parameters from the reflected path's parameters to estimate accurate ToF~\cite{802.11az} and AoA~\cite{spotfi}.
So, \name's first key idea is to obfuscate the direct path such that the first of the reflected paths appears to be the shortest traveled path. This would in turn lead to inaccurate AoA and ToF estimates, degrading their localization accuracy to be of the order of a few meters as shown in Figure~\ref{fig:main-intro-fig}(b).

\begin{figure}
    \centering
    \includegraphics[width=0.7\linewidth]{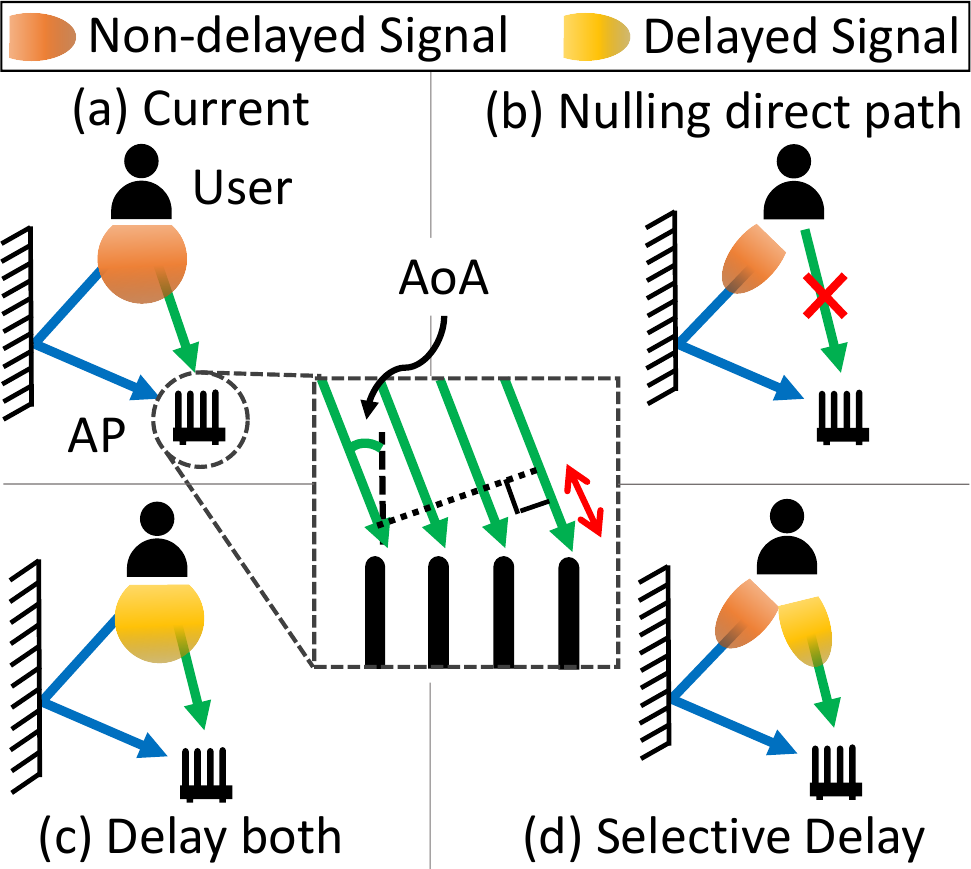}
    \caption{(a) Direct path (green) arrives earlier at the AP due to shorter path length (b) Nulling the direct path obfuscates the AoA, but reduces signal strength (c) Naively delaying the paths does not change the relative arrival times, (d) \name selectively delays the direct path, breaking the assumption in (a). Inset: relative phase accrued from path length difference due to AoA across receive antennas}
    \label{fig:intro}
\end{figure}

% Para 6 -- A dumb way to do this is nulling but reduces the communication link, so we firstly precode such that the ap does not perceive the correct direct path and also does not remove any paths thus maintatining communication link.
A simple way to make the algorithm estimate the reflected path to be the shortest path is to simply remove the direct path as shown in Figure~\ref{fig:intro}(b). This removes one of the dominant paths for connectivity, reducing the throughput of the user connection. Thus we cannot simply remove the direct path signal. We also note that for a given location of the user and the Wi-Fi AP, we cannot change the directions or AoAs of the received signal as that depends on the geometry of the environment that is out of our control or would need additional hardware deployments that compromise deployability of \name's design. So, in \name we identify that the only way we can let the receiver perceive the reflected path to be the shortest traveled path is if we can "delay" the signal along the direct-path direction.

% Para 7 -- The second challenge, is that delaying one-path creates delays or ghost peaks across all the paths like done in mirage, so we need to ensure that the direct path and multipath precoding matrices are orthogonal and that we add delay to only the direct path's precoding vector.
Since the multiple reflections (multipath) in the environment, including the direct path and the reflected paths, have a common source, naive beamforming as performed by the authors in MIRAGE~\cite{ayyalasomayajula2023users} that applies a delay to the transmitted signal, unfortunately, would delay all the signals in all directions by the same amount as shown in Figure~\ref{fig:intro}(c). Here \name makes a key observation that most of the modern user device relies on MIMO communication and thus have multiple antennas for transmission. This would imply that the user device can beamform~\cite{beam_forming_doc} the signal independently in each of the signal departure directions or angle-of-departures (AoD). Thus, in \name we design an optimized precoding matrix that is conditioned to beamform the transmitted signal while ensuring that both:
(a) The beamforming along the direct path with additional delay is orthogonal to the reflected path's transmitted directions.
(b) The beamforming performed for the reflected paths without additional delays is orthogonal to the newly delayed reflected path's transmitted directions.

% Para 8 -- Finally, we address a few real world problems of consistency of the direct path and multiple collaborative APs, by adding random delays and obfuscating one AP at a time across all the environment AP respectively.
Finally, \name also addresses a key real-world problem to make the system robust to a range of attacks -- Since, direct-path is always more consistent across time than reflected path~\cite{spotfi}, an attacker who has knowledge of \name can simply cluster packets across time as done in ~\cite{spotfi} to get the most consistent path. We ensure to overcome this by randomizing the obfuscation delay added to the direct path, so even the direct path is no longer consistent over time. 
% (b) An attacker can also compromise multiple APs, to locate the user. We demonstrate \name such that it can obfuscate each AP the user is connected to.

\name's obfuscation is tested on commercially available Wi-Fi radios~\cite{warp}, across 30 locations and over 1000 packets in realistic settings. \name's design degrades the accuracy for fundamental predictions of AoA and ToF that are used by the state-of-the-art localization algorithms~\cite{vasisht2016decimeter, soltanaghaei2018multipath,liu2021rfloc,shoudha2023wifi,ma2019wifi,802.11az}. So we compare the accuracy degradation provided by \name when using two state-of-the-art localization systems -- SpotFi~\cite{spotfi} (relies fundamentally on AoA) and UbiLocate~\cite{pizarro2021accurate} (relies fundamentally on ToF). Through these experiments, we demonstrate the efficiency of \name on Wi-Fi user's location protection. Our experimental results show the localization error degrades by $\sim6\times$ at the median with localization errors as high as $1.33$ m in a $4$m$\times4$m space, AoA estimation error degrades by $20\times$ at the median, and ToF estimation error degrades by $\sim2.5\times$ at the median after \name. 

\noindent\textbf{Our Contributions:}
\begin{itemize}[leftmargin=*]
    \item \name's defense obfuscates all of the current decimeter accurate localization systems by up to $6\times$, and ensures at best a meter-level or worse accuracy.
    \item \name's defense does not require any hardware or firmware changes, and is Wi-Fi standard compliant.
    \item An attacker cannot counter \name despite of their knowledge of \name's implementation, owing to the randomization of obfuscation delay.
\end{itemize}

% {\todo{we need the contribution section}}
% !TEX root = paper.tex

\section{Threat Model}\label{sec:attack}
We formally describe the attack model and boil down to the requirements and the basic assumptions we make in designing \name's defense model.\looseness-1
% {\color{red}ToDo: 1. discuss the situation of the enterprise network, the client's capability with multi-antenna, and AP's capability. 2. add 2.4ghz and 6ghz discussion for the sake of scope. 3. the scope of the threat model, e.g., if the attacker knows this defense 4. cots phone has linear array or not, which needs to be clarify. 5. seems that the reviewers need to know uplink/downlink obfus, and what knowledge we need to have for uplink obfuscation.}

\noindent \textbf{Attack scenario.} We consider the scenario in the Wi-Fi network that is widely deployed in enterprise buildings (e.g., office buildings, malls, or other public venues), where the Wi-Fi APs are compromised or deliberately deployed by the attacker to communicate with the Wi-Fi users and reveal their location without their consent. We label this as the ``attacking'' APs. The Wi-Fi APs and users can communicate with the standard Wi-Fi protocols on any frequency band. To do steal the Wi-Fi user's location privacy, the attacker can hack into a single-AP localization system, where the Wi-Fi AP can derive the angle-of-arrival (AoA) and time-of-flight (ToF) measurements for localization. The attacker can hack into multiple APs that can collaborate with each other to leverage the angle of arrival (AoA) to localize the Wi-Fi user. Compromising the Wi-Fi user's location privacy.  Finally, we should be able protect the Wi-Fi user's location even though the attacker knows our defense implementation.

To this end, we seek to protect the billions of daily Wi-Fi users today whose location privacy is at stake. To provide this privacy protection, we explore how we can obfuscate the AoA and ToF measured by the attacking AP via a software-only solution. %{\color{red}ToDo: add the discussion on when there is no reflection.}

\noindent \textbf{Multipath indoor wireless environment.} In typical indoor wireless environments such as offices, rooms, cafes, malls, etc. there are different kinds of reflectors in the environment such as walls, chairs, desks, etc. The multipath signal propagation in the indoor wireless environment is very common as indicated in ~\cite{zwick2002stochastic}.

\noindent \textbf{Attacking AP.} 
%Concerningly, both of these parameters can be determined without the user's consent, as this sensing can be a byproduct of the communication between the AP and the user. From a single received packet, 
The attacking AP can estimate AoA using numerous state of the art systems~\cite{spotfi, xie2018md, arraytrack, rapmusic,soltanaghaei2018multipath} from a single received packet. To estimate the ToF of the user from the attacking AP, WiPeep~\cite{wipeep} accurately measures the round trip time using the protocol-enforced ACK message generated by the user's Wi-Fi device. Through these systems, an attacking AP has two components to determine the user's location with decimeter-grade accuracy, and the user is rendered defenseless against the localization attack.

\noindent \textbf{Wi-Fi user.} We consider the Wi-Fi users who can estimate the downlink wireless channel and leverage it to obfuscate his/her location through precoding the uplink data streams. The Wi-Fi user can be instrumented with an antenna array for MIMO communication in the wireless environment.

\noindent \textbf{Scope.} To successfully deploy \name, we make two assumptions that often hold true in real-world settings:
\begin{itemize}[leftmargin=*] 
    \item The user device and the AP are equipped with multi-antenna, linear antenna arrays. 
    \item The wireless channel is reciprocal, i.e. the downlink (AP to user device) and uplink (user device to AP) channels are equivalent. 
\end{itemize}

\section{Primer on Wireless Localization}\label{sec:primer:localization}
The threat model provides the scope of the problem we are solving. In this section, we will first provide a quick primer on the necessary concepts needed to localize a user with a single AP. Second, we will demonstrate on how naive solutions to obfuscate the user location deteriorates the signal strength and communication throughput. 

\begin{figure}
    \centering
    \includegraphics[width=0.85\linewidth]{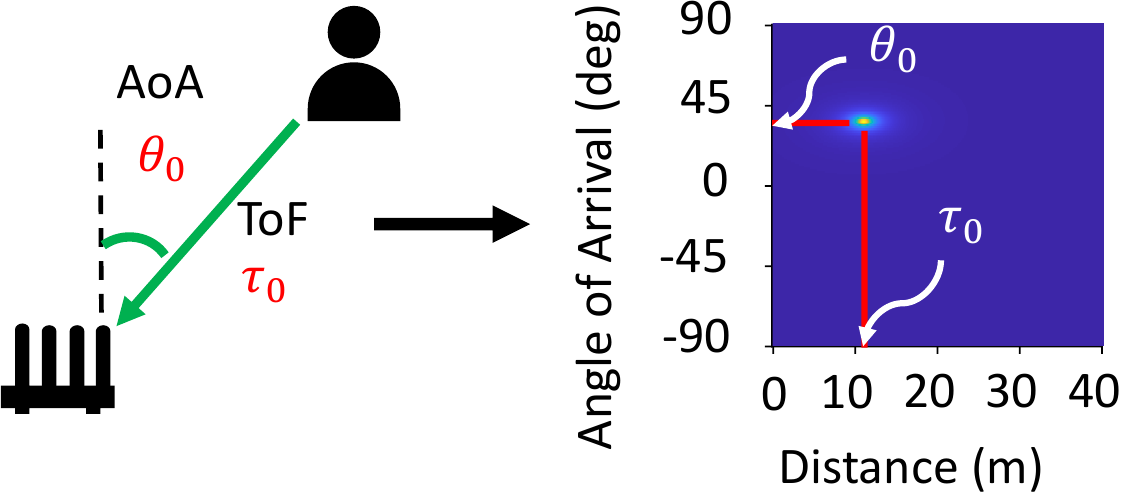}
    \caption{Wi-Fi-based localization with estimated AoA and ToF. After the Wi-Fi AP obtains the AoA $\theta_0$ and ToF $\tau_0$, the user's location is pinpointed. The right figure shows the profile with the SpotFi algorithm for AoA and ToF identification.}
    \label{fig:primer:localization}
\end{figure}

\subsection{Indoor localization}%\label{sec:attack-model}

To localize the user in the Wi-Fi infrastructure, we consider the simple scenario where there are only direct path signals between the Wi-Fi AP and the user as shown in ~\cref{fig:primer:localization}. The Wi-Fi AP is trying to figure out the Wi-Fi user's coordinates (i.e., location) $\vec{p}_\mathrm{user}$ within his/her own coordinates $\vec{p}_\mathrm{AP}$. As a result, Wi-Fi AP can estimate the user's (x, y) coordinates by
$$ \vec{p}_\mathrm{user} = \vec{p}_\mathrm{AP} + c\tau_0*[\cos(\theta_0) \quad \sin(\theta_0)]^T, $$
where $\vec{p}_\mathrm{AP}$ is the location of the Wi-Fi AP, $\tau_0$ and $\theta_0$ are the estimated ToF and AoA in the AP's coordinates. Basically, AoA can tell in which direction the Wi-Fi user is and ToF can tell how far away the Wi-Fi user is. The Wi-Fi AP can run the SpotFi~\cite{spotfi} algorithm and further identify $\theta_0$ and the UbiLocate~\cite{pizarro2021accurate} algorithm to estimate the ToF $\tau_0$. Therefore, a single Wi-Fi AP can simply localize the Wi-Fi user with the above equation. The following illustrates how the attacker can obtain AoA and ToF for the Wi-Fi user's localization. 

\noindent \textbf{ToF estimation.} To estimate the distance between the attacker and the Wi-Fi user, we can estimate the round-trip time (i.e., $2\times$ToF) of the packets transmitted between the attacker and the Wi-Fi user. Then, the distance between the attacker and the Wi-Fi user is the multiplication of ToF and light speed. To do so, we can leverage the Fine Time Measurement (FTM) protocol~\cite{ftm_secure} to obtain the ToF of the packets transmitted between the attacker and the Wi-Fi user. To further tune the ToF estimation, we can leverage the wireless channel state information (CSI). As a signal travels over the air, it accrues additional phases depending on its frequency. A higher frequency signal will accrue a larger phase. We can leverage this fact to fine-tune the ToF measurements further. Specifically, a Wi-Fi signal is transmitted over multiple linearly spaced frequencies. Measuring the relative phase accumulation across these frequencies provides an estimate of the ToF of the signal. 

\noindent \textbf{AoA estimation.} After an attacking AP estimates the ToF, the attacker can figure out the distance between the attacker and the Wi-Fi user. To localize the Wi-Fi user, the attacker still needs to know in which direction the Wi-Fi user is. To this end, an attacking AP with multiple antennas can measure the phase difference of the incoming Wi-Fi signal between pairs of antennas to estimate the AoA. For simplicity, imagine two antennas spaced $d_\mathrm{sep}$ apart. The phase difference accrued between these two antennas is related to the AoA $\theta_0$ as 
$$\Delta \Phi = \exp(-\iota 2\pi \frac{1}{\lambda} d_\mathrm{sep} \sin(\theta_0)),$$ 
where $\lambda$ is the wavelength corresponding to the signal's center frequency and $d_{sep}$ is the antenna separation on the receiver. This additional phase difference is a consequence of the incremental distance signals travel to each of the receive antennas, as visualized in \cref{fig:intro}. Most of the existing localization algorithms rely on this fundamental effect to measure the AoA. 

These fundamental effects allow most localization algorithms to estimate the WiFi signal's path parameters. A visualization of the SpotFi transform, which estimates the ToF and AoA of the signal, is shown \cref{fig:primer:localization}. This profile will be used later as a visual aid to describe \name's obfuscation. 
\looseness-5

\begin{figure*}[ht]
    \centering
    \includegraphics[width=\linewidth]{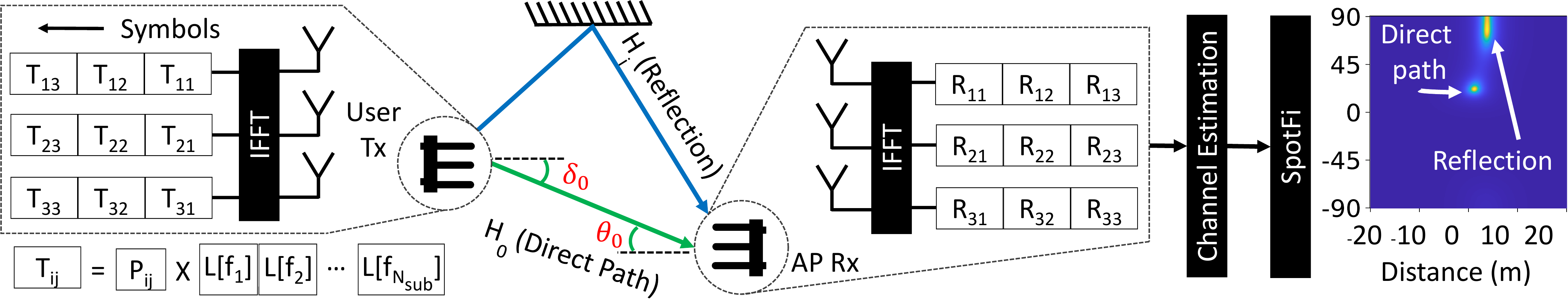}
    \caption{\textbf{Wireless Channel based User Localization:} The user transmits a known precoded signal across $3$ time symbols from the $3$ antennas. The signals travel via multiple paths (green, blue) to the AP. The channel can be estimated based on the known transmitted signal. SpotFi~\cite{spotfi} is applied to segregate the paths across the angle and distance domain.}
    \label{fig:channel-model}
\end{figure*}

%\subsection{Path towards protecting privacy}\label{sec:attack-algo}
%To estimate the user's location, we need two measurements -- the time of flight and the angle of arrival. In WiPeep, the authors proposed a potential solution to protect against ToF measurements. However, their solution requires hardware changes and is only applicable to future Wi-Fi devices. 

\subsection{Nulling to the direct path} \label{sec:nulling}

To disable the attacking AP from leveraging the direct path signals for localization, a straightforward idea is to null the direct path. As a result, the attacking AP has to use multipath for localization, thereby, the location privacy is protected. To do so, we can create a nulling precoder $P_\mathrm{obf} = \vec{v_n}\vec{v_n}^H$, where $\vec{v_n}$ is the nulling vector in the direct path signal's null space, i.e. $\vec{v_d}^H \vec{v_n} = 0$. However, nulling the direct path will reduce the signal strength at the receiver and degrade the communication throughput at the attacking AP. Through experimental evaluation, we find the RSSI degrades by $6$ dB when the direct path is nulled. Instead, \name preserves all the paths and chooses to delay the direct path \textit{selectively} beyond the reflected paths to create ambiguity in AoA measurement. \name effectively increasing user localization error to over 3 meters while reducing signal strength by only $3$ dB. The impacts of nulling and \name's clever precoding are further characterized in \cref{sec:tput}. In the remaining sections, we will present how \name applies the appropriate precoding without affecting the channel capacity and obfuscates the channel unbeknownst to the attacking AP.

\section{Designing Software-level Obfuscation}\label{sec:design}

We consider a single AP localization system to be one of the most deployable localization systems for Wi-Fi user localization. Most enterprise settings often deploy APs in an optimal fashion where only a single AP is visible to a Wi-Fi user. To comply with these realistic constraints, advancements have been made in AoA (Angle of Arrival) and ToF (Time of Flight) estimation to leverage these ubiquitous Wi-Fi signals for localization.

\subsection{Intuition on location obfuscation}

In this section, we show how we can protect the Wi-Fi user's location privacy by obfuscating the wireless signals. Let's first compute the phases accrued at both the transmit and receive array of linearly spaced antennas due to the additional relative distance the signal travels:
\begin{equation}
\tiny
    \vec{v_0}=\begin{bmatrix}
                1\\ 
                e^{-\iota2\pi d_\mathrm{sep}\sin\delta_0 / \lambda}\\ 
                e^{-\iota2\pi (2d_\mathrm{sep})\sin\delta_0 / \lambda}\\ 
                \vdots\\
                e^{-\iota2\pi (N_{TX}-1)d_\mathrm{sep}\sin\delta_0 / \lambda}
              \end{bmatrix} \, 
  \vec{u_0}=\begin{bmatrix}
                1\\ 
                e^{-\iota2\pi d_\mathrm{sep}\sin\theta_0 / \lambda}\\ 
                e^{-\iota2\pi (2d_\mathrm{sep})\sin\theta_0 / \lambda}\\ 
                \vdots\\
                e^{-\iota2\pi (N_{RX}-1)d_\mathrm{sep}\sin\theta_0 / \lambda}
              \end{bmatrix}
\end{equation}
If a transmitter steers power in the $\delta_0$ direction, the phases it must apply on its $N_{TX}$ antennas are given by $\vec{v}_0$. Similarly, if the receiver captures a signal coming from angle $\theta_0$, then the phases given in $\vec{u}_0$ will be accrued at the $N_{RX}$ antennas. These angles have been visualized in \cref{fig:channel-model}. For an $N_{RX} \times N_{TX}$ wireless system, the Wi-Fi signal's channel for a given single path, which is transmitted over $N_{sub}$ equally spaced frequency subcarriers, can be written as
$$ H_0(f_i) = a_0 \vec{u}_0 \vec{v}_0^H \exp(\iota 2\pi f_i \tau_0) \in \mathcal{C}^{N_{RX} \times N_{TX}},$$
where $a_0$ and $\tau_0$ are the direct path signal amplitude and ToF. As discussed in the introduction, we see the delays due to ToF appear as an additional phase dependent on the frequency of the $i^{th}$ subcarrier. 

However, the Wi-Fi signal reflections in the environment create ambiguities in AoA measurements. An example is shown in \cref{fig:channel-model}, with the direct path depicted by the green arrow and the reflection(s) by the blue arrow. If there are $M$ more paths that the signal takes to reach the receiver, the channel becomes

\begin{align}\label{eq:chan-ap}
    H(f_i)  = H_d(f_i) + \underbrace{\sum_{p=1}^M a_p \vec{u}_p \vec{v}_p^H \exp(\iota j 2 \pi f_i \tau_p)}_{\text{Reflections: } H_r}
\end{align}
where  $H_d(f_i)=a_d e^{-j 2 \pi f_i \tau_0} \vec{u}_d \vec{v}_d^H$ is the direct-path channel and $\vec{u}_d$ and $\vec{v}_d$ represent the steering vectors at the receiver and transmitter for direct path signals. Similarly, $\vec{u}_p$ and $\vec{v}_p$ represent the steering vectors at the receiver and transmitter for reflected path signals.

The reflections additively combine with the direct path channel defined previously and create ambiguous AoA measurements, $\theta_p$. 
However, the state-of-the-art localization algorithms assume the direct path, a straight line between the user and attacker AP, travels the least distance. The direct path is distinguished by either estimating the ToF through successive optimization~\cite{xie2018md, rapmusic} or via special transforms~\cite{spotfi, bloc, soltanaghaei2018multipath}, and choosing the path with the least ToF, $\tau_p$. An example of the Spotfi transform, which segregates the wireless channel according to the AoA and ToF. Here, it is apparent that the direct path travels the least distance over the air and arrives first. The direct path's angle can be discerned from the y-axis. In \name, we seek to exploit this core assumption that the direct path always arrives the earliest to confuse the attacking AP.

The following section will first discuss how the channel is estimated at the attacking AP (\cref{sec:chan-est}). In \cref{sec:design-bf}-\ref{sec:add-refs}, we build on the nulling idea explored previously to selectively delay only the direct path while preserving the communication throughput. 
\begin{figure*}[ht]
    \centering
    \includegraphics[width=0.9\linewidth]{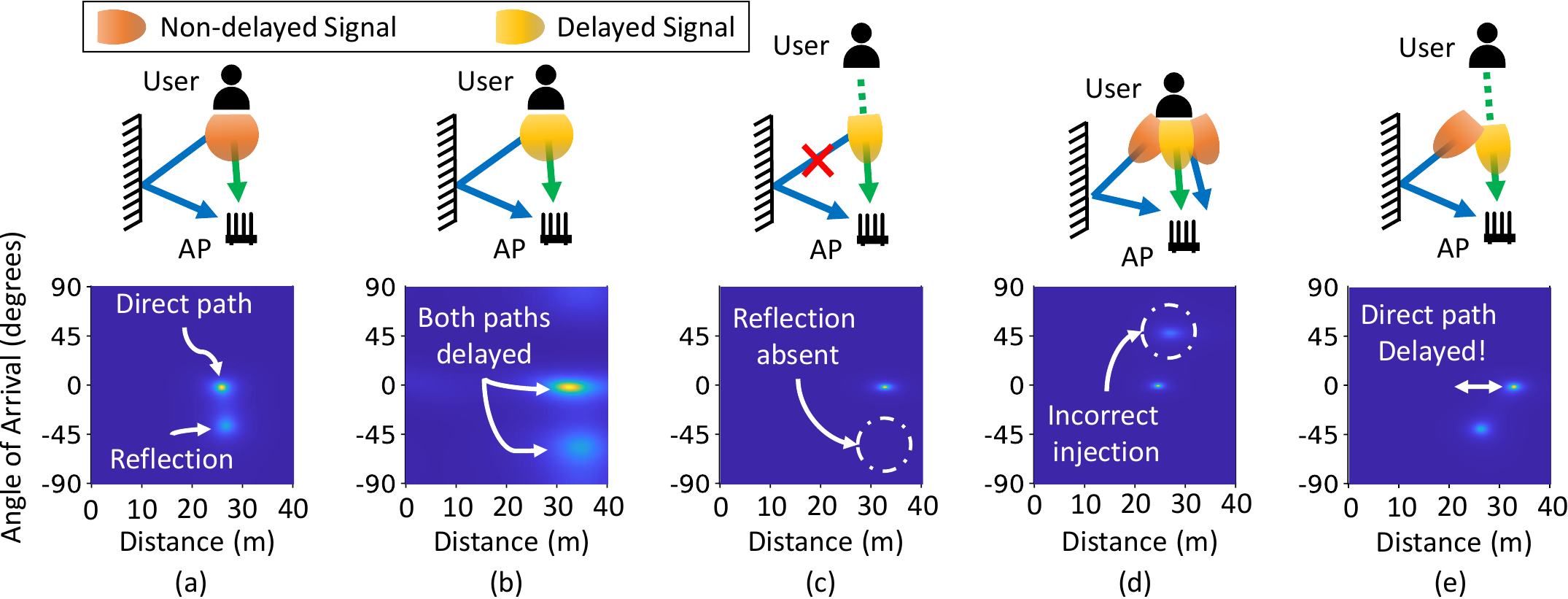}
    \caption{(a) SpotFi profile of channel simulated assuming 4 RX antennas, 4 TX antennas, and 40 MHz bandwidth; (b) applying delay and beamforming towards direct path, both paths are delayed; (c) applying delay and beamforming \textit{solely} towards direct path and nulling elsewhere, removes reflections; (d) re-injecting reflections naively distorts the profile; (e) re-injecting reflections while aware of direct-path obfuscates channel.}
    \label{fig:precoding-ablation}
\end{figure*}

\subsection{Channel estimation at the attacking AP}\label{sec:chan-est}

In the standardized MIMO system, the wireless channel between each pair of transmitting and receiving antennas needs to be estimated as a necessary prerequisite to decoding the data. A known preamble sequence is used to estimate this channel. This sequence is transmitted across $N_{TX}$ symbols across all the transmit antenna.
% , as shown in \cref{fig:channel-model}. 
The received signal for frequency $f_i$ across the $N_{RX}$ antennas and the $N_{TX}$ symbols is given by
\begin{align*}
    & R[f_i] = H(f_i) T[f_i], \quad H, R \in \mathcal{C}^{N_{RX} \times N_{TX}}, T \in \mathcal{C}^{N_{TX} \times N_{TX}} \\
    & \implies H(f_i) = R[f_i] T[f_i]^{-1}
    % & R[f_i] = H(f_i) T[f_i] \implies H(f_i) = R[f_i] T[f_i]^{-1} \\
    % & where \quad H(f_i) \in \mathcal{C}^{N_{TX} \times N_{RX}}, and \quad R[f_i], T[f_i] \in \mathcal{C}^{N_{TX} \times N_{TX}} \\
\end{align*}

To ensure the channel can be reliably estimated, an invertible $T$ is chosen. It is simply constructed across all frequencies by considering a preamble symbol $\vec{L} \in \mathcal{C}^{N_{sub}}$, and an invertible precoding matrix $P[f_i] \in \mathcal{C}^{N_{TX} \times N_{TX}}$. Hence, the user can transmit the data $T[f_i] = P[f_i] \vec{L}[i]$ that is precoded by the precoding matrix, and the channel is estimated via 
$$H(f_i) = \frac{R[f_i]}{\vec{L}[i]} P[f_i]^{-1}$$ 
which will be used by the attacker to localize the user. As a result, the user can manipulate the precoding matrix of  $P[f_i]$ to obfuscate his/her location.

Since the channel is estimated at the receiver, a user does not have any impact on the way the attacker measures the channel or how it traverses the environment. Hence, many past works~\cite{yao2018aegis, staat2022irshield} seeking to obfuscate user locations deploy additional hardware in the environment to change the wireless channel explicitly, making their widescale implementation challenging. However, the user does have control over the precoding ($P[f_i]$) they can apply on the preamble sequence at a software level. Hence, without loss of generality, if the attacking AP expects an identity matrix to be applied as precoding, but the user instead applies a precoding matrix $P_\mathrm{obf}[f_i]$, the attacking AP will measure an incorrect obfuscated channel 
\begin{align}
    H(f_i)_\mathrm{obf} &= H(f_i) P_\mathrm{obf}[f_i] \nonumber \\
                        &= a_d e^{-j 2 \pi f_i \tau_0} \vec{u}_d \vec{v}_d^H P_\mathrm{obf}[f_i] + H_r(f_i) P_\mathrm{obf}[f_i] \label{eq:chan_obf} 
\end{align}
Next, we will illustrate how we can manipulate the precoding matrix such that the user's location is obfuscated and the communication throughput is maintained. 

% \begin{table}[]
%     \small
%     \begin{tabular}{|c|c|c|cccc|}
%     \hline
%     \multirow{2}{*}{} &
%       \multirow{2}{*}{No obf.} &
%       \multirow{2}{*}{Null} &
%       \multicolumn{4}{c|}{Preserve paths and delay } \\ \cline{4-7} 
%      &  &  & \multicolumn{1}{c|}{0 (m)} & \multicolumn{1}{c|}{20 (m)} & \multicolumn{1}{c|}{30 (m)} & 40 (m) \\ \hline
%     AoA error &
%       $0^\circ$ &
%       $62^\circ$ &
%       \multicolumn{1}{c|}{$0^\circ$} &
%       \multicolumn{1}{c|}{$58^\circ$} &
%       \multicolumn{1}{c|}{$61^\circ$} &
%       $53^\circ$ \\ \hline
%     \begin{tabular}[c]{@{}c@{}}RSSI\\ (dBm)\end{tabular} &
%       -65 &
%       -71 &
%       \multicolumn{1}{c|}{-64} &
%       \multicolumn{1}{c|}{-64} &
%       \multicolumn{1}{c|}{-64} &
%       -62 \\ \hline
%     \end{tabular}
%     \caption{Received signal strength for different obfuscation techniques}
%     \label{table:aoa_rssi}
% \end{table}

\subsection{Selective beamforming and controlled delay}\label{sec:design-bf}

\begin{figure}[t]
    \centering
    \includegraphics[width=0.5\linewidth]{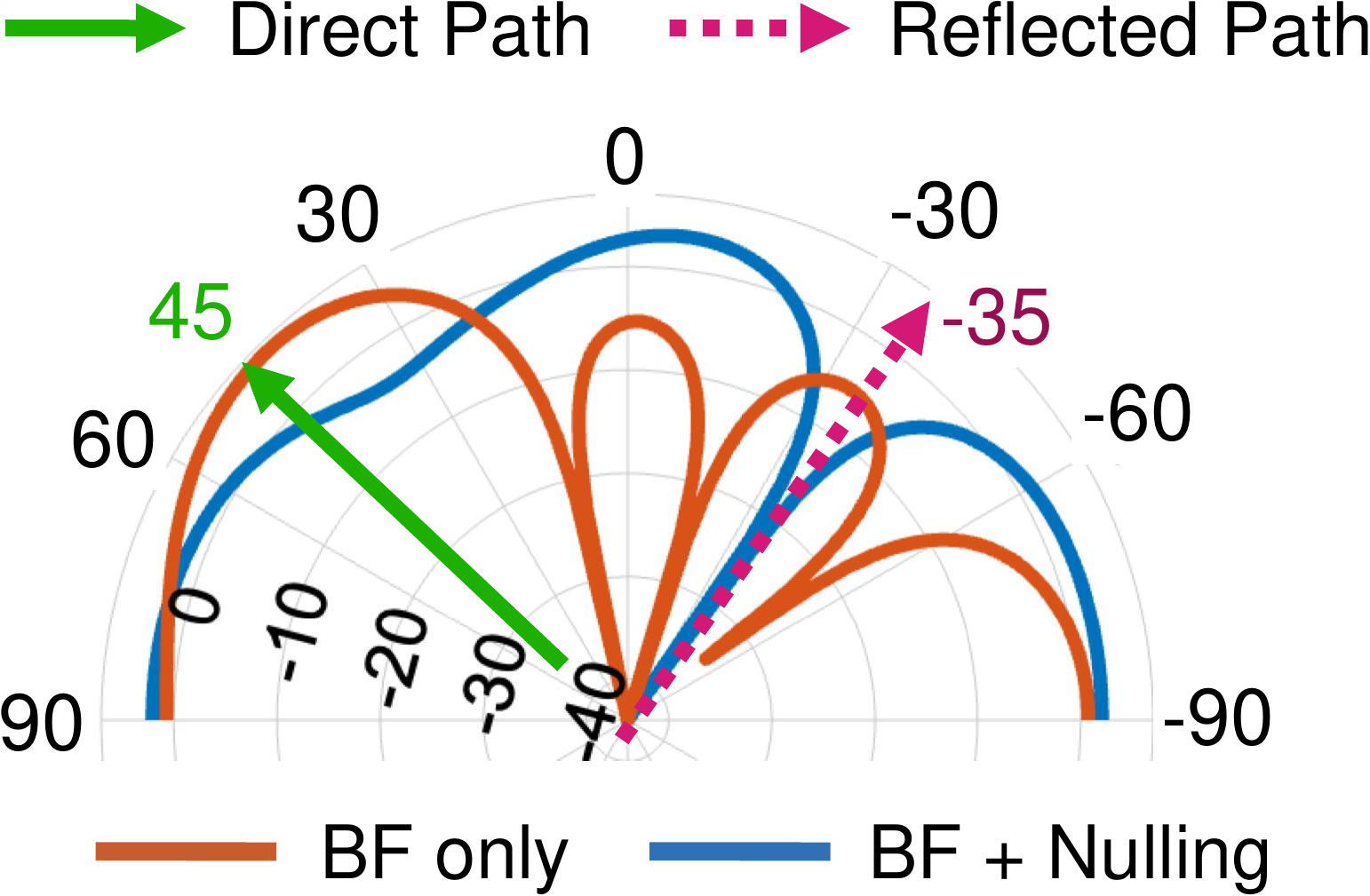}
    \caption{Antenna patterns show power directed towards different directions. Without nulling, power is directed towards direct path and reflections, delaying both paths equally}
    \label{fig:ant-pattern}
\end{figure}

%{\color{red} ToDo: 1. add a description of the symbols before using it. 2. channel estimation process should be more detailed 3. put some results (e.g., throughput is not compromised without design) here to make the flow smooth and understandable}
\name's core deliverable is to increase $\tau_0$, the direct path's time of flight, to a number larger than that of the reflections. It should achieve this by minimally impacting the rest of the channel. The simplest way to achieve this is to beamform solely in the direction of the direct path, and apply the necessary delay to the channel. In this case, $P_\mathrm{obf} = \vec{v_d}\vec{v_d}^H \exp(j 2\pi f_i \tau_\mathrm{delay})$. This effectively delays the direct path but at the detrimental cost of delaying the multipath signals by similar amounts. This is visually depicted for an example beamforming pattern in \cref{fig:ant-pattern}. It is clear that the direct path beamformer (orange), leaks power in the direction of the reflections (pink). This can lead to the reflections accruing the same delays as the direct paths. Hence leaving the relative ToFs between the various paths unchanged and allowing the attacking AP to discern the direct path as the earliest arriving signal. The impact of this is shown in \cref{fig:precoding-ablation}(b), wherein both the direct path and reflected paths are delayed by similar amounts when compared to the original channel in \cref{fig:precoding-ablation}(a). Hence, our first insight is to apply delay via an appropriate precoder without impacting the reflections in $H_r$.

A simple and effective precoder hence lies in the null space of $H_r$, which ensures power is not directed in the directions of the reflections. Since $H_r$ is the channel devoid of direct path, it is not full rank, implying the existence of at least one such null vector $\vec{n}_r$. We can then construct $P_\mathrm{obf} = \vec{n}_r \vec{v}_d^H \exp(j 2\pi f_i \tau_\mathrm{delay})$. This is shown by the blue beamformer in \cref{fig:ant-pattern}. This beamformer does not direct power in the direction of the reflection, allowing \name to solely delay the direct path. However, the nulling vector does not align with the direct path direction perfectly, and can create incorrect delays applied to the channel. Specifically, we have $\vec{v}_d^H\vec{n}_r = \alpha \exp(j\phi)$ which needs to be explicitly compensated for within the precoder, giving us
$$P_\mathrm{obf}[f_i] = \frac{1}{\alpha}\vec{n}_r \vec{v}_d^H \exp(j 2\pi f_i \tau_\mathrm{delay} - \phi)$$
As a result, the direct-path channel will be properly delayed without affecting the reflected paths.

%\todo{cite mimo paper, talking about spatial multiplexing}

Unfortunately, this precoder, which lies in the null space of the reflected paths, entirely suppresses these reflections which are crucial to obfuscate the channel, i.e., $H_r (f_i) P_\mathrm{obf}[f_i] = 0$. In the traditional MIMO communication systems~\cite{tse2005fundamentals}, the communication throughput is enhanced through the multiplexing gain by leveraging the antenna array at the transmitter and receiver, as the multiple antennas can further explore the space diversity in the physical wireless environment over constructive and destructive signals addition. As a result, this additionally reduces the channel capacity by reducing the spatial multiplexing gain, the ability to support multiple parallel streams of data communication, which negatively affects the overall throughput of the MIMO system leading to lower data rates. This impact is visualized in \cref{fig:precoding-ablation}(c).

\subsection{Adding back the reflections} \label{sec:add-refs}
%{\color{red} ToDo: 1. clearly specify the distortion introduced by adding hr 2. distinction between the authorized access network and attacker. are they the same? why attacker cannot get encryption key?}

We can supplement the precoder with a beamformer directing power along the reflected paths to bring these reflections back. However, naively adding back $H_r$ creates distortions to the channel as visualized in \cref{fig:precoding-ablation}(d). This is a consequence of non-zero power directed in the direction of the direct path.  
Clearly, adding back the reflections is trickier than it seems at first glance, as any changes we now make to the precoding impact the direct path delays as well. 

We can circumvent the impact on the direct path by simply choosing to reconstruct the multipath components using the null vectors of the direct path steering vector $v_d$. Specifically, given the null space of $\vec{v}_d^H$ as $N \in \mathcal{C}^{N_{TX} \times (N_{TX}-1)}$, the goal is to supplement the above precoder with an additional component which is closest to identity, 
\begin{align*}
    P_\mathrm{supplement}[f_i] = NA \approx I \implies A = N^H (N N^H)^{-1} 
\end{align*}
This supplementary precoder will not impact the direct path component of the channel but will reintroduce the multipath most effectively. Hence, \name's final precoder is
\begin{align}
P_\mathrm{obf}[f_i] = \frac{1}{\alpha}\vec{n}_r \vec{v}_d^H \exp(j 2\pi f_i \tau_\mathrm{delay} - \phi) + N N^H (N N^H)^{-1} \label{eq:prec-bf-delay}
\end{align}
In this way, even though we add back the reflections, the direct path signals will not be affected. The impact of applying this precoder on our channel is shown in \cref{fig:precoding-ablation}(e).

\subsection{Generalizing to Multi-AP}\label{sec:design-multiap}
%\todo{add by Wei Sun}

In the above discussion, we have shown the feasibility and theory of obfuscating the AoA and ToF measurements with our innovative precoding weight. For the sake of simplicity, we illustrate it with the attack settings where there is only one Wi-Fi AP leveraging AoA and ToF to localize the Wi-Fi user. This is a quite popular setting in the real-world wireless communication environment. However, in a typical enterprise network, multiple Wi-Fi APs can be deployed to provide wide communication coverage for Wi-Fi users. As such, the attacker can coordinate multiple APs and leverage the AoAs derived from them to triangulate the Wi-Fi user. Therefore, we only need to obfuscate AoA information derived at the Wi-Fi APs to trick them with incorrect user locations. Since the Wi-Fi user's localization in a multi-AP scenario highly depends on the coordination of the AoAs derived at multiple APs, our \name can still protect the Wi-Fi user's location privacy even though we can only obfuscate one Wi-Fi AP's location at one time. We demonstrate this experimentally in Section~\ref{sec:results}.

\section{Enhancing location obfuscation through randomization}

\begin{figure}
    \centering
    \includegraphics[width=\linewidth]{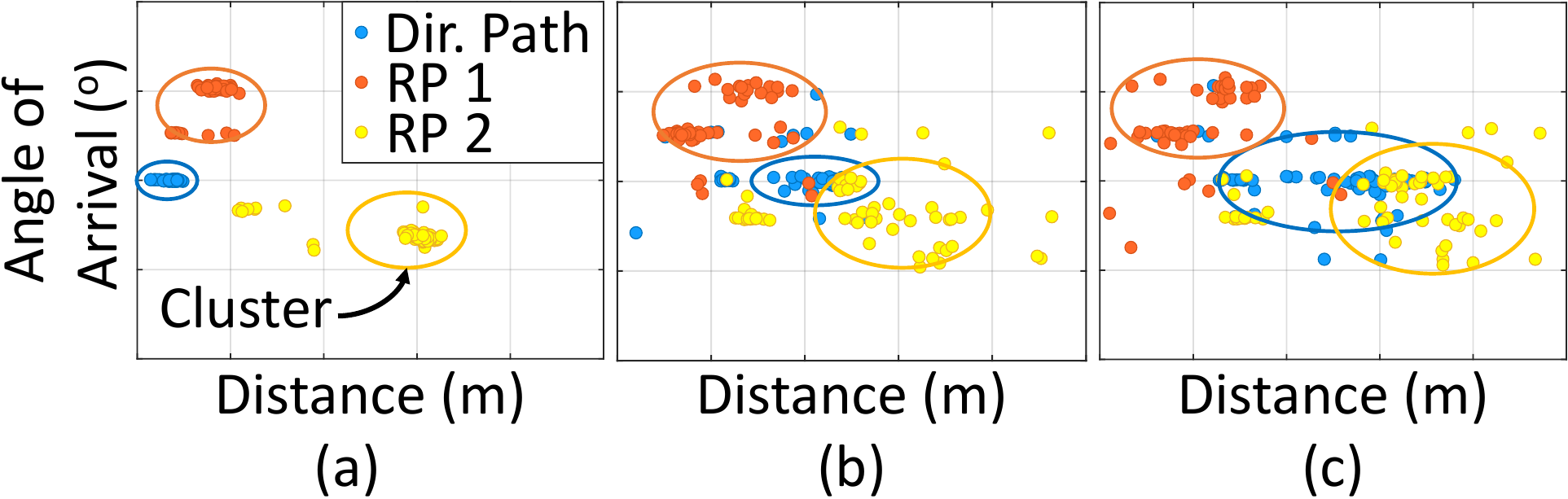}
    \caption{ (a) Direct path peaks (in blue) cluster more tightly as compared to reflected paths (RP1, in orange, and RP2, in yellow) and can be used to evade \name's obfuscation. (b) Adding the constant delay for obfuscation, while the cluster indicating the direct-path signals is still concentrated, results in easy direct-path signal detection. (c) Adding randomized delay to the obfuscation over consecutive packets broadens the direct path clustering.}
    \label{fig:cluster}
\end{figure}

The previous section discussed the precoding weight \name that applies to obfuscate the channel. This precoding weight selectively delays the signal transmission along the direct path to ensure that the first signal arriving at the attacking AP is a reflection. This creates ambiguity for the attacker in estimating the direct path's AoA and ToF, hence obfuscating the user's location. However, it is important to ensure that the attacker cannot break our location obfuscation through reversed engineering. To this end, we strengthen the obfuscation to an attacker who is trying to circumvent \name.  

%\subsection{Obfuscating despite knowing defense}\label{sec:mirage-known}
%{\color{red}ToDo: need more details and security analysis rigorously. I think we should either put it at the discussion or do micro-benchmark or simulation here.}

Specifically, a clever attacker may try to circumvent the defenses provided by this system. A simple way to circumvent this attack is by averaging packets across multiple time steps. As the user moves through space, often the angle of arrival of reflections in the environment changes more rapidly as opposed to the direct path links. The direct path is hence more stable across these packets. If the attacker collects Wi-Fi packets over several seconds and averages the channel over these packets, they can effectively reduce the impact of reflections in the environment while strengthening the effects of the direct path. This is often done in Wi-Fi localization systems~\cite{arraytrack} to further improve localization accuracy. By reducing the effects of reflections in the perceived channel, \name cannot ``hide'' the direct path amongst reflections, impairing our obfuscation capabilities. A simple insight can protect against this attack -- impart a similar randomness to the direct path as well. We can do this by effectively applying random delays, $\tau_\mathrm{delay}$ to the direct path. This will ensure the direct path representation is also weakened during the averaging operation, affecting the attacker's ability to measure the direct path's AoA and ToF.

Therefore, we can add random delay to the direct path signals to disable the attacker from identifying the direct path signals through averaging. As a demonstration, we did an experiment to show the feasibility of enhancing location obfuscation through randomization. As shown in Fig.~\ref{fig:cluster}(a), one direct path signal and two reflected path signals are exhibited in the indoor environment with concentrated clusters. When we add constant delay to the direct path signals as shown in Fig.~\ref{fig:cluster}(b), we can clearly see the cluster indicating the direct path signals are shifting. However, it is still easy to be identified through the clustering and the cluster is concentrated. As a result, this breaks our location obfuscation with constant delay. However, when we apply the random delay to the direct path signals as shown in Fig.~\ref{fig:cluster}(c), we can see that the cluster indicating the direct path signals becomes more sparse and diverse, which can further confuse the attacker for direct-path signal identification. As such, our location obfuscation approach is reliable even though the attacker is aware of this obfuscation technique.

% !TEX root = paper.tex
\section{Implementation}\label{sec:imp}

%This section briefly reviews the deployment challenges of implementing \name, and the infrastructure and user-side hardware system setup to thoroughly test the ideas presented in this work. For some microbenchmarks, we will also utilize the simulator to quantify the design choices made in developing \name. This section will also provide further details about this simulator. 

\noindent \textbf{Estimating path parameters:}
With the precoder mathematically defined in \cref{sec:design-bf}, we need to estimate the core variables that are key to creating this precoder on the user's device. To compute the precoding, we need knowledge of the uplink channel, the wireless channel when the user transmits to the AP, collected at the attacking AP. We can rely on channel feedback, wherein the AP informs the user of the uplink channel and is a standard part of MIMO communication. However, it would defeat our efforts if the attacking AP could access an unobfuscated channel. Instead, we assume that the channel is reciprocal and note that the user's Wi-Fi device will listen to the AP-discovery packet before connecting to an AP. The hermitian of the downlink channel measured on this packet can be assumed to be the uplink channel we need~\cite{tse2005fundamentals}. 

We can now compute the precoding with access to the pseudo-uplink channel. The two components of the precoder require the direct path channel and the channel comprising the reflections. We have to tease apart that direct path and reflected path components. We can reliably do this via mD-Track~\cite{xie2018md}. 

\begin{figure}
    \centering
    \includegraphics[width=\linewidth]{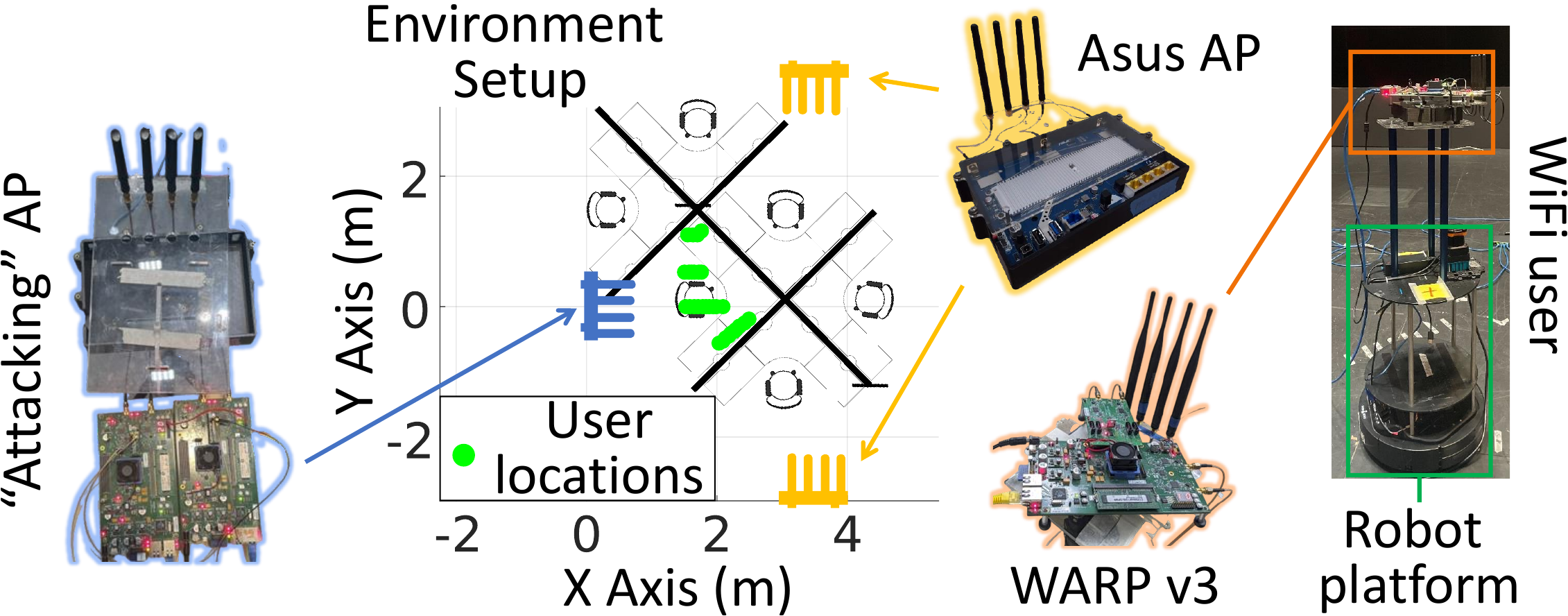}
    \caption{%\todo{edit caption} 
    Experimental Setup: On the left is the Dual-WARP setup that we use as the attacker AP in blue color, and the rightmost image shows the User WARP device that is mounted on a robot that collects ground-truth location for the user device. The green dots are the user locations \name is tested at, and the AP is at origin (0,0). Two ASUS AP-enabled APs in yellow color are at (3.5, 3.56) and (3.5, -2.45) respectively.}
    \label{fig:ap:warps}
\end{figure}

\noindent \textbf{Infrastructure:} We use a software-defined radio (WARP~\cite{warp}) as the Wi-Fi AP instrumented with four antennas working at the 5GHz Wi-Fi band as shown by the blue highlight in  Fig.~\ref{fig:ap:warps}. The Wi-Fi user obfuscates this ``Attacking'' AP for location privacy protection. An ASUS AP~\cite{asus} is co-deployed with this WARP to obtain the ToF of packets exchanged between the Wi-Fi AP and the user via the Ubilocate~\cite{pizarro2021accurate} algorithm. UbiLocate conducts FTM packet exchanges to coarsely estimate the ToF (labeled as ``Vanilla'' FTM). These coarse measurements are fine-tuned with the channel measurements collected at the ASUS AP as discussed in \cref{sec:primer:localization}, and are referred to as ``Enhanced'' ToF measurements. To characterize the obfuscation performance of ToF estimation with UbiLocate, we provide the obfuscated channel measurements made at the WARP instead of those collected at the ASUS AP. As such, the Wi-Fi AP can leverage the derived AoA and ToF to localize the Wi-Fi user.

Although in most scenarios, only a single AP is visible to a user in enterprise settings (as explored in \cref{sec:eval-ap-assc}), there are some cases where multiple APs can collaborate simultaneously to localize a user. To test \name's obfuscation in these ``multi-AP'' scenarios, we deploy two additional ASUS Wi-Fi APs as shown in Fig.~\ref{fig:ap:warps}. Subsequently, we can use a total of three Wi-Fi APs to triangulate the Wi-Fi user with the derived AoA measurements. These additional Wi-Fi AP's will mainly leverage the SpotFi~\cite{spotfi} algorithm to create the AoA-ToF profile, which can be used to estimate the received signal angle of arrival (AoA). Note, however, that \name cannot simultaneously obfuscate the AoA estimation at these additional APs. We will instead discover that an incorrect AoA measurement at one of the three APs corrupts the triangulation algorithm and degrades location accuracy. 

\noindent \textbf{User:} Similar to the radios on the infrastructure side, we use a software-defined radio (WARP) as the Wi-Fi user instrumented with four antennas, as shown in the right of the Fig.~\ref{fig:ap:warps}. We co-deploy an Asus AP on the user side as well to perform FTM exchanges with the Asus AP, which is co-deployed with the infrastructure WARP. These radios are deployed on a robot to emulate the Wi-Fi user's movement.

\noindent \textbf{Experimental Setup:} To measure the performance of our system, we conducted experiments in an indoor area of the typical enterprise building. Specifically, the Wi-Fi user moves around in the indoor environment on a robot, as shown in Fig.~\ref{fig:ap:warps}, and the Wi-Fi infrastructure localizing this user is static in the indoor environment.  The Wi-Fi AP and user can communicate with each other over the 5.18 GHz center frequency, corresponding to Wi-Fi channel 36. Note that we can always configure three Wi-Fi APs and one Wi-Fi user to different center frequencies without affecting our obfuscation performance. During the communication, the Wi-Fi AP may localize the Wi-Fi user with the AoA estimated with channel measurements from WARPs and ``enhanced'' ToF estimated from FTM measurements made by ASUS APs and channel measurements made from the ASUS or WARP. Specifically, our experimental setup can enable single AP-based Wi-Fi user localization by using SpotFi to derive the AoA measurement and UbiLocate to derive the ToF measurement. Alternatively, the three Wi-Fi APs can collaborate with each other to triangulate the Wi-Fi via AoA measurements alone.

\noindent \textbf{Simulator Setup:} We use a simulator to quantify the impact of our design choices on the obfuscation performance. Via simulations, we implement some of the naive design choices mentioned in \cref{sec:design} and showcase their detrimental impact on obfuscation performance. We build the simulator in MATLAB and implement the channel transformations we expect to see over the air. The channel models are implemented as described in MDTrack~\cite{xie2018md} and in \cref{sec:design}. We randomly simulate the addition of reflections, their angle of arrival at the receiver, and their angle of departure at the transmitter and realistically scale the powers of the paths based on real-world measurements from our experimental setup. We apply $10$ dB of AWGN noise to simulate standard indoor environments. The open-source SpotFi and Ubilocate systems are utilized for AoA and ToF estimation.

\section{Evaluation}\label{sec:results}

Our evaluation consists of end-to-end system-level tests and microbenchmarks. We evaluate the localization accuracy with and without \name to demonstrate the end-to-end location obfuscation errors in the dynamic indoor environment in \cref{sec:end-end}. Then, we show the effect of location obfuscation on the wireless communication performance in \cref{sec:tput}. We report the robustness of our location obfuscation approach and system in the dynamic indoor environment in \cref{sec:mb}.

We consider a simple case study to motivate the need for \name further. In a typical enterprise network such as a company building, the administrator may stealthily localize the office staff to monitor their time in the cubicles and characterize their productivity~\cite{lopez2017human}. Specifically, the staff's mobile phone can connect with the Wi-Fi AP operated by the office administrator. With state-of-the-art localization techniques, the staff can be localized within a few tens of centimeters, which suffices to confidently localize them within their cubicle. As a result, the office administrator could invade their staff's privacy and police their performance. To this end, the staff can leverage our technique to obfuscate his/her location. The following sections delve deeper into the \name's obfuscation performances and how they relate to this simple case study.

\begin{figure*}[ht]
\begin{subfigure}{0.24\textwidth}
\centering
\includegraphics[width=\textwidth]{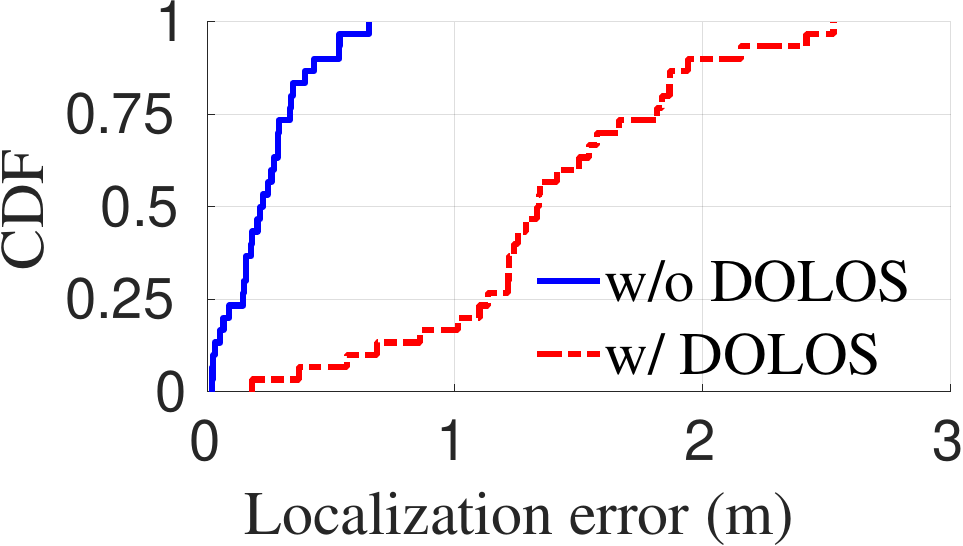}
\caption{Single-AP with AoA+FTM}%\todo{changed} {\color{red} checked}}
\end{subfigure}
\begin{subfigure}{0.24\textwidth}
\centering
\includegraphics[width=\textwidth]{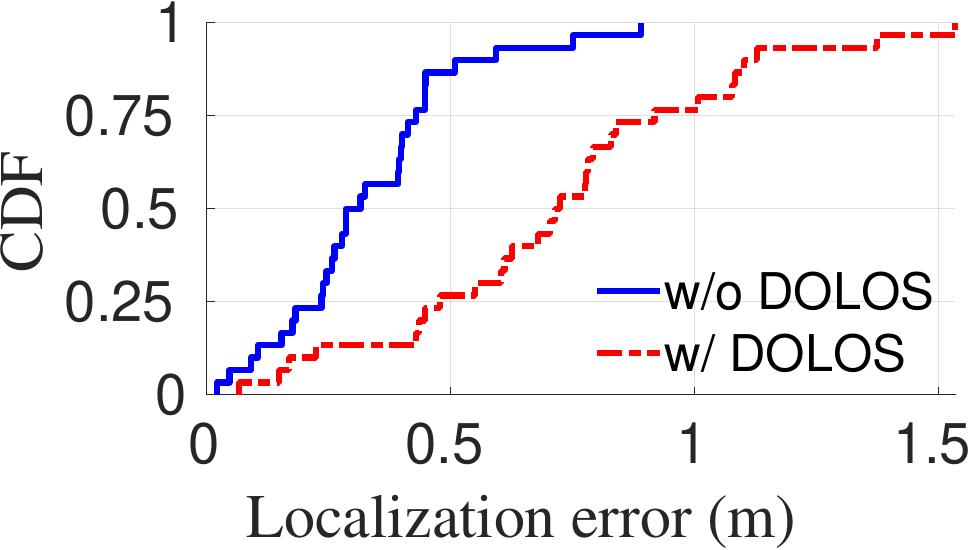}
\caption{Triangulation with AoAs}
\end{subfigure}
\begin{subfigure}{0.24\textwidth}
\centering
\includegraphics[width=\textwidth]{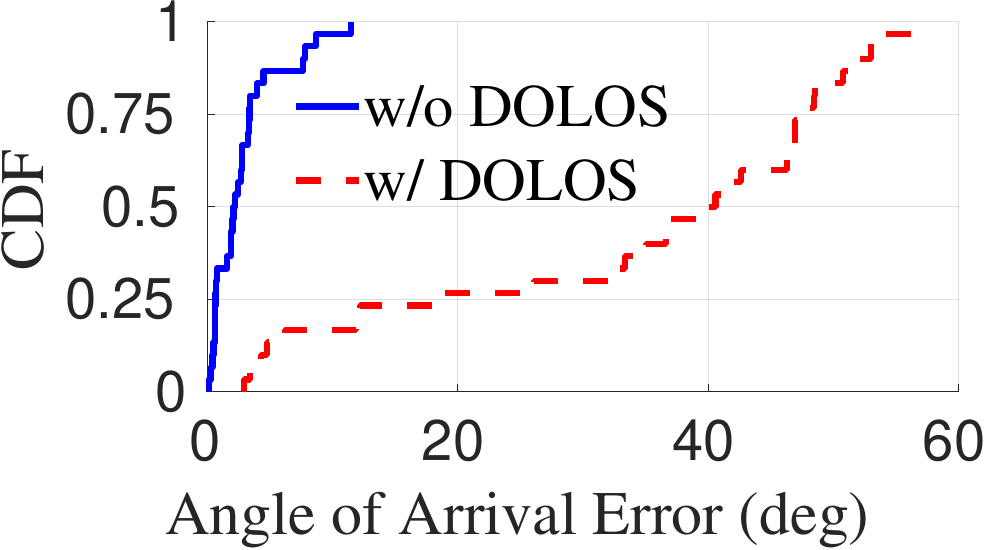}
\caption{AoA estimation}
\end{subfigure}
\begin{subfigure}{0.24\textwidth}
\centering
\includegraphics[width=\textwidth]{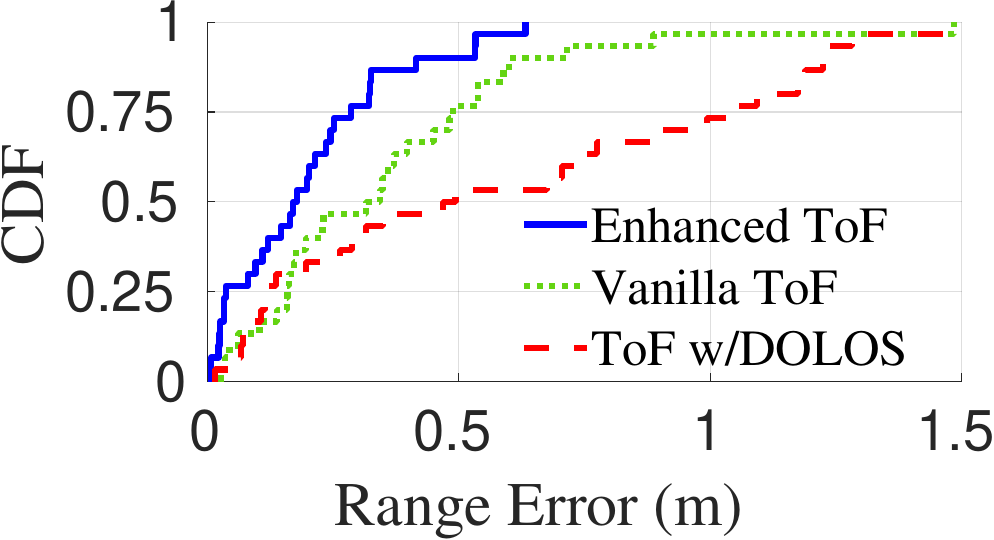}
\caption{Range estimation}
\end{subfigure}
\caption{%\todo{fix caption, {\color{red} checked}}
End-to-end localization error with and without \name in the (a) single-AP scenario and (b) multi-AP scenario through triangulation. (c) AoA estimation error with and without \name. (c) ToF estimation error with ``vanilla'' FTM, ``enhanced'' ToF, and ToF w/\name.}
\label{fig:end-to-end}
\end{figure*}

\begin{figure*}
    \centering
    \includegraphics[width=\linewidth]{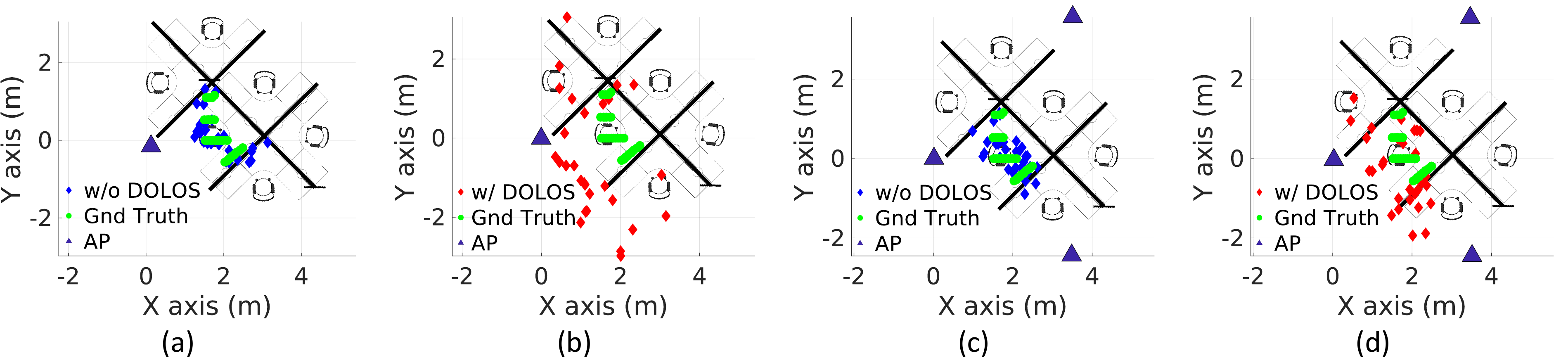}
    \caption{%\todo{{\color{red} checked}}  
    Localization with AoA and FTM in the single-AP scenario (a) without \name enabled and (b) with \name enabled. Triangulation with AoAs derived from three APs (c) without \name enabled and (d) with \name activated.}
    \label{fig:scatter}
\end{figure*}

\subsection{End-to-end System Evaluation}~\label{sec:end-end}

In this section, we will thoroughly evaluate \name's obfuscation on the localization performance using the two state-of-the-art localization systems -- SpotFi~\cite{spotfi} and UbiLocate~\cite{pizarro2021accurate}. Specifically, we will show the localization, AoA estimation, and ToF estimation errors of these algorithms with and without \name's obfuscation applied. 

\subsubsection{Location obfuscation in single-AP scenario}
The localization error is the difference between the estimated location and the ground-truth location of the Wi-Fi user. We conduct experiments in an office building to show the localization errors of \name. The Wi-Fi user's location is estimated through the uplink obfuscated channels at the Wi-Fi AP through the super-resolution algorithm Spotfi to estimate the AoA. We rely on UbiLocate to provide us with decimeter-accurate ranging estimates. With both the AoA and ToF we can estimate the user's location. 
 
\cref{fig:end-to-end}(a) shows the CDF of the localization error while using these state-of-the-art localization systems. Specifically, we use SpotFi to obtain the AoA estimation and UbiLocate to derive the ToF measurement in the single-AP localization scenario. We then apply \name's obfuscation before transmitting the packet. As we can see, the median localization error is around $0.215$ m before obfuscation. However, with \name activated, the localization error increases $\sim6\times$, with the median localization error around $1.33$ m. This implies with \name's obfuscation enabled, even state-of-art localization systems' accuracy reduces to that of RSSI-based systems~\cite{ez}, indicating the significance of \name's algorithm to obfuscate the Wi-Fi user's location.

\noindent\textbf{Casestudy:} \cref{fig:scatter} (a) and (b) show the estimated Wi-Fi user's locations using AoA and FTM in the single-AP scenario with and without \name. In the single-AP scenario, an employer can localize an employee within their cubicle with $86.6\%$ accuracy. Encouragingly, we can see that the accuracy rate degrades to $16.6\%$ when \name is enabled, helping preserve the employee's location privacy.

\subsubsection{Location obfuscation in multi-AP scenario}

In the multi-AP scenario, multiple APs can collaborate with each other to improve localization accuracy. To show the efficiency of our \name on degrading the localization performance in the multi-AP scenario, we deploy three APs in the indoor environment to localize the Wi-Fi user through triangulation. 

\cref{fig:end-to-end}(b) shows the CDF of the localization error while using the state-of-the-art localization systems to triangulate the Wi-Fi user. Specifically, we use the SpotFi algorithm to derive the AoAs at the multiple APs to triangulate the Wi-Fi user. We then apply \name to obfuscate the AoA estimation on one of the APs. As we can see, the median localization error is around $0.28$ m before the obfuscation. However, when the \name is enabled, the localization error increases $\sim2.5\times$ with a median localization error of around $0.71$ m. This indicates that the state-of-the-art localization systems degrade to the RSSI-based localization systems, resulting in the location privacy protection.

\noindent\textbf{Casestudy} \cref{fig:scatter} (c) shows the localization performance without enabling \name. As we can see, the employee can be accurately localized within the cubicle $93.3\%$ of the time. This is because the multiple APs can collaborate with each other to triangulate the Wi-Fi user accurately. However, with \name enabled at the Wi-Fi user, the classification accuracy significantly deteriorates to $53.3\%$. This is shown in \cref{fig:scatter} (d). Even though \name can only obfuscate one Wi-Fi AP, it can significantly break the collaboration among APs, resulting in efficient location privacy protection. %\todo{talk about this later?} We can discuss more on the impact of the number of APs on the obfuscation performance in our discussion section. 

\subsubsection{Degrading AoA estimation}
Since \name attempts to obfuscate the angle-of-arrival information of the Wi-Fi user, it's crucial to demonstrate the performance of \name on AoA obfuscation. To do so, we plan to show the AoA errors after applying our \name algorithm. 

To localize the Wi-Fi user, the Wi-Fi AP usually leverages the direct-path AoA and ToF for Wi-Fi user localization. Here, we report the AoA estimation error at the Wi-Fi AP with and without obfuscation. The AoA estimation error is defined as the difference between the ground-truth AoA and the estimated AoA with a super-resolution localization algorithm (SpotFi).

\cref{fig:end-to-end}(c) shows the CDF of the AoA errors while using SpotFi without obfuscating the channel and with \name's obfuscation activated. As we can see, the median and $90^\mathrm{th}$ percentile AoA error without \name is around $2.0^\circ$ and $7.8^\circ$ , respectively. However, this AoA prediction with SpotFi severely degrades by $20\times$ at the median and $6.6\times$ at the $90^\mathrm{th}$ percentile once \name obfuscates the channel. This is because \name can obfuscate the direct-path AoA and confuse the Wi-Fi AP to use the random multipath for Wi-Fi user localization, resulting in high AoA estimation errors. Furthermore, this AoA obfuscation mainly results in the location obfuscation to the state-of-the-art localization systems.

\subsubsection{Degrading ranging estimation}
As previously explored, FTM measurements can be further refined by relying on the phase measurement across subcarriers to re-estimate the time-of-flight of the signal. This improves the ranging estimates between the user's devices and the AP. This improvement is seen in \cref{fig:end-to-end}(d), where the ``enhanced-ToF'' measurements median accuracy is 0.18 m, whereas the ``vanilla-FTM'' performs $~2\times$ worse, with median accuracy of 0.344 m. Consequently, these enhancements will be incorporated within the 802.11az ranging scheme~\cite{802.11az}. 

Interestingly, the same phase measurements used to fine-tune the ToF measurements are impacted when \name's obfuscation is applied. Specifically, the enhanced ToF measurements are obtained by measuring the ToF of the direct path signal. But, \name creates ambiguity between the direct path and multipath, subsequently leading to the AP measuring a reflection's ToF instead. Since these reflections travel a longer distance than the direct path, the ToF is over-estimated, leading to errors in ranging estimates. From \cref{fig:end-to-end}(d), we can see that the range estimated by estimated ToF with \name's obfuscation turned on are $~2.5 \times$ worse, with a median ranging error of $0.49 m$.

\begin{figure*}
    \begin{subfigure}{0.33\linewidth}
        \includegraphics[width=0.95\linewidth]{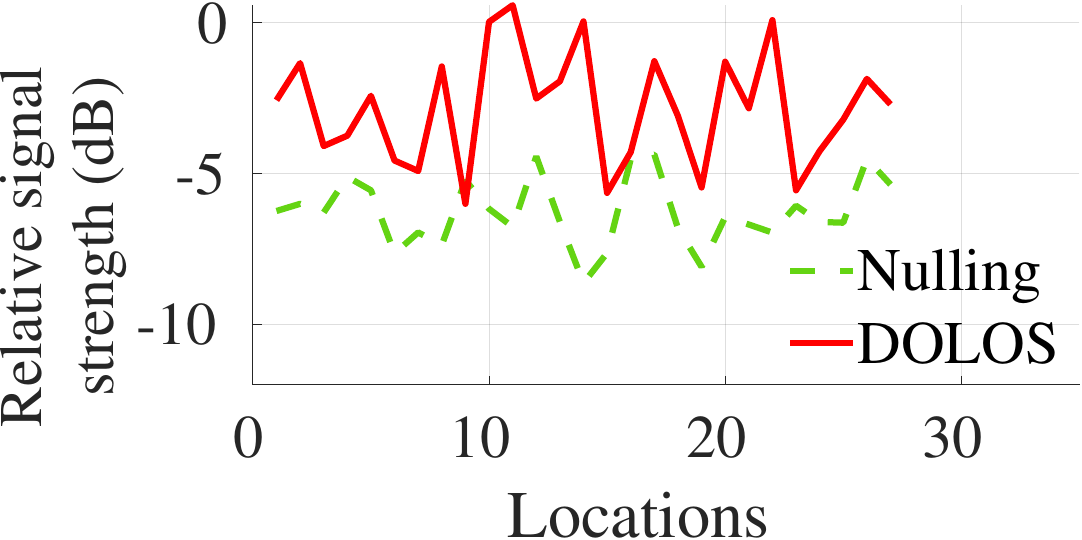}
        \caption{Maintaining the throughput performance}
    \end{subfigure}
    \begin{subfigure}{0.33\linewidth}
        \includegraphics[width=\linewidth]{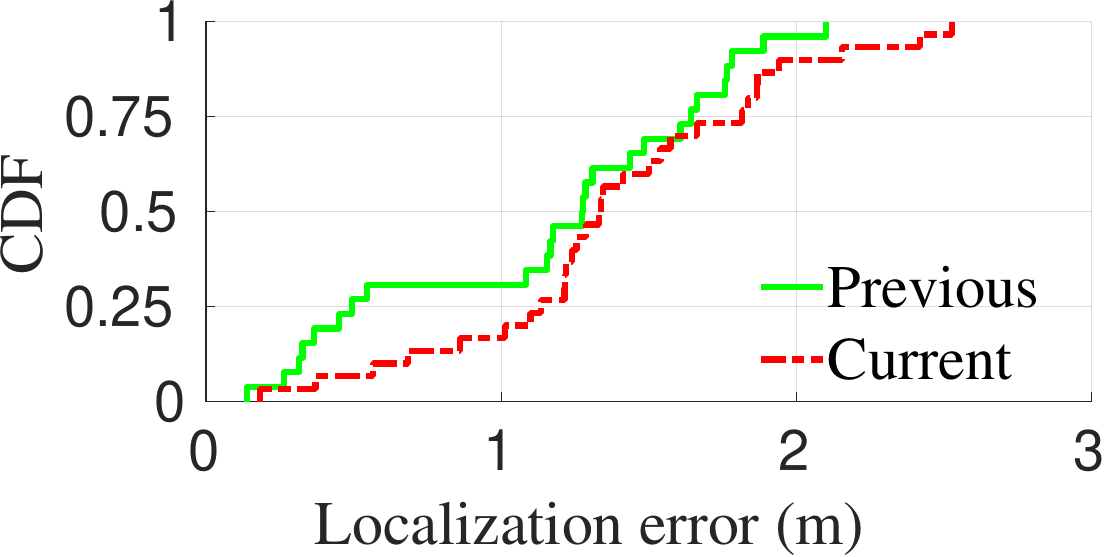}
        \caption{%\todo{changed} {\color{red}checked} 
        Localization error under motion}
    \end{subfigure}
    \begin{subfigure}{0.33\linewidth}
        \includegraphics[width=\linewidth]{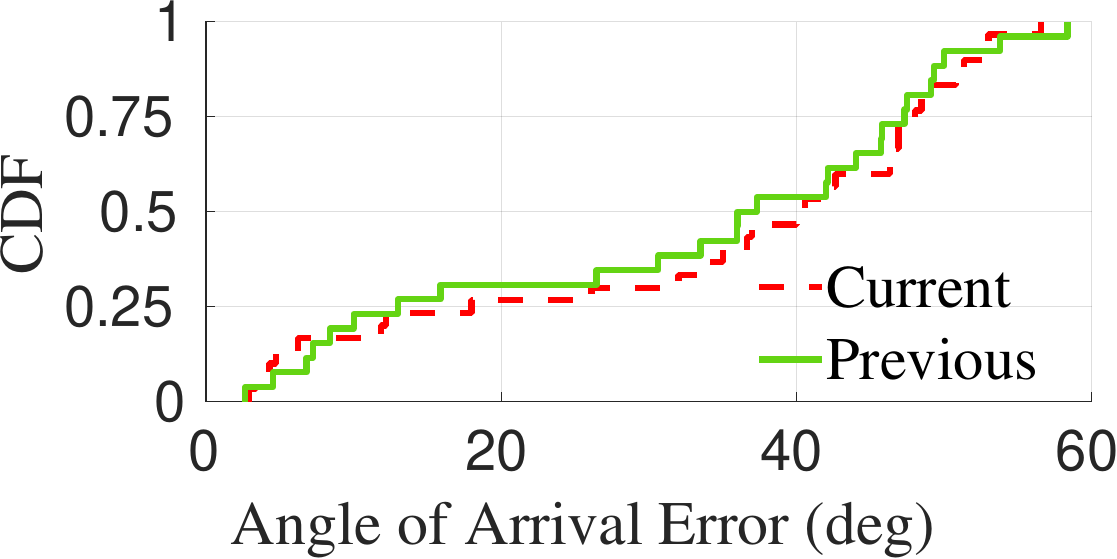}
        \caption{AoA estimation error under motion}
    \end{subfigure}
    \caption{(a) Relative signal strength across different Wi-Fi user's locations when nulling to the direct path and \name are applied. (b) Single-AP localization error and (c) AoA estimation error when \name leverages the current and past packets to generate the precoding matrix for obfuscation. }
    \label{fig:end-to-end2}
\end{figure*}

%{\color{blue} discuss the correlation of throughput and rssi} {\color{red} checked}
\subsection{\name's Impact on throughput}~\label{sec:tput}

Now that we have established \name's ability to effectively obfuscate the user locations via obfuscating both the AoA and ToF measurements, this obfuscation must not impact the throughput at the user. An effective metric to measure channel throughput is via the signal strength measurements when applying the various obfuscation schemes. This evaluation is also employed by the prior work~\cite{lin2011random,eletreby2017empowering,gadre2020frequency,gollakota2009interference} to measure the channel throughput. Specifically, we care about the drop in signal strength due to our obfuscation compared to an unobfuscated packet transmission. In \cref{fig:end-to-end2}(a) we showcase this relative signal strength when \name and a simple nulling scheme are applied for the hardware setup in \cref{sec:end-end}. We argued in \cref{sec:design} that simply nulling towards the direct path will provide a reliable way to obfuscate the user's location. However, this nulling severely impacts the signal strength and the channel throughput. We can see an average of $~3 dB$ additional drop in signal strength when using this naive nulling scheme compared to using \name for obfuscation. \name's precoding creates an average drop of $~3.4$ dB in signal strength compared to when the transmission is unobfuscated.

\subsection{Microbenchmarks}~\label{sec:mb}

This section will delve deeper into the design choices made in \name and further evaluate the robustness of the obfuscation scheme. Specifically, we will look at how \name performs under moving scenarios with access to stale channel estimates, understand the impact of multipath-diversity in the environment, and the need for the different ideas presented in \cref{sec:design}.

\subsubsection{Testing Robustness under motion}

To measure the robustness of \name under motion, we do the experiments by leveraging the channels measured when the Wi-Fi user moves. These measured channels will be used by the Wi-Fi user to create the precoding discussed in \cref{sec:design} and to transmit a modified packet to the Wi-Fi AP, such that the AP cannot accurately identify the direct-path AoA and ToF for localization. However, the channel measured under motion might be from a past time step for a different user location. \name's obfuscation must continue to work in the scenarios where a ``stale'' packet is used to estimate the precoding. 

In \cref{fig:end-to-end2}(b), we explore \name's performance when generating the precoding with this stale packet in the \textit{single-AP scenario}. For this evaluation, the precoding is estimated based on the packet collected during the previous timestep and applied at the current timestep. Between these time steps, the user has moved an average of ~$7$ cm. As we can see, the median localization error when using the previous packet is around $1.27$ m, and $90\mathrm{th}$ percentile localization error is around $1.78$ m, providing a similar obfuscation performance compared to using the current packet for precoding. Similarly, \cref{fig:end-to-end2}(c) shows the AoA estimation error when using this stale precoding vector for location obfuscation. As we can see, the median AoA estimation error is around 36.71 degrees, and the 90th percentile AoA estimation error is $~50$ degree, which is close to using the current precoding vector for location obfuscation. This is because the wireless channel is quasi-static, and our location obfuscation approach is resilient to the dynamic wireless environment.

\begin{figure}
    \centering
    \includegraphics[width=0.9\linewidth]{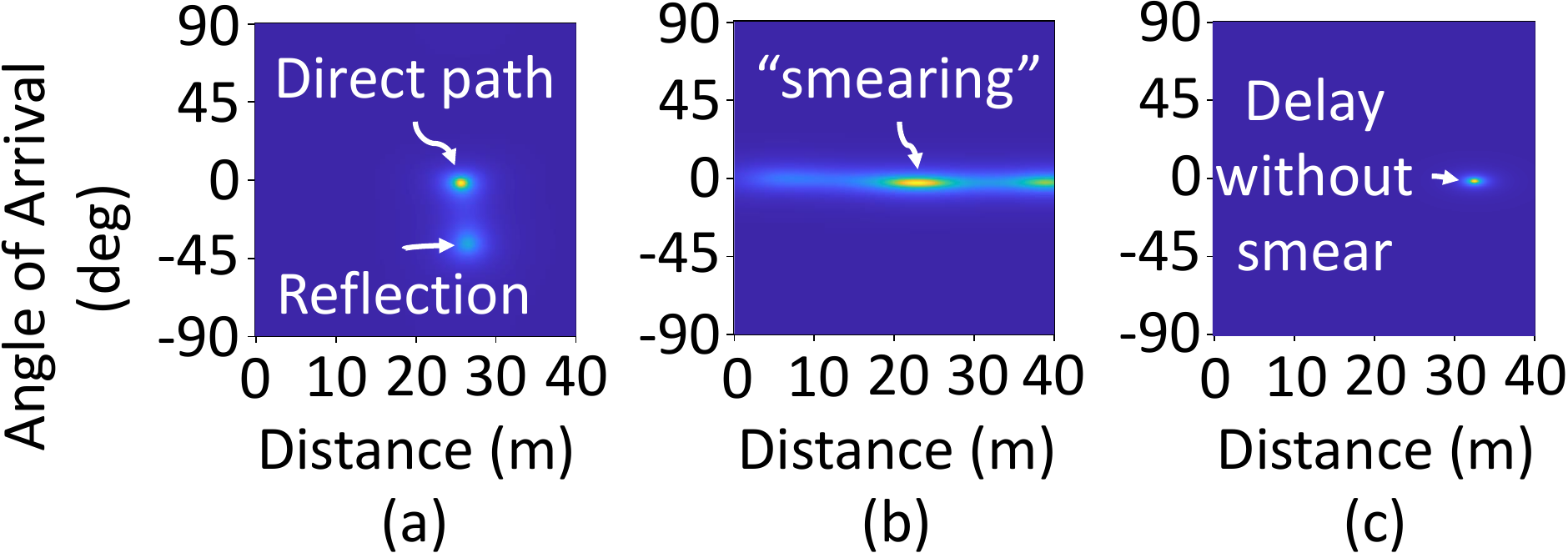}
    \caption{(a) Original channel. (b) Channel without compensation shows the residual direct-path channel. (c) Channel with compensation and the residual direct path channel is compensated.}
    \label{fig:comp-need}
\end{figure}

\begin{figure*}
    \begin{subfigure}{0.32\linewidth}
        \includegraphics[width=\linewidth]{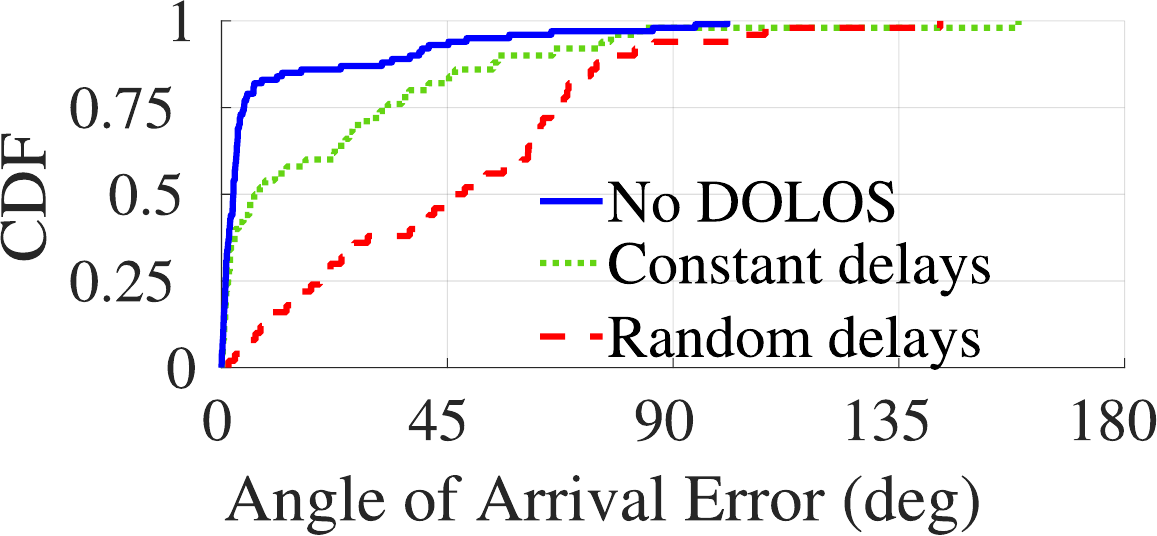}    
        \caption{}
    \end{subfigure}
    \begin{subfigure}{0.31\linewidth}
        \includegraphics[width=\linewidth]{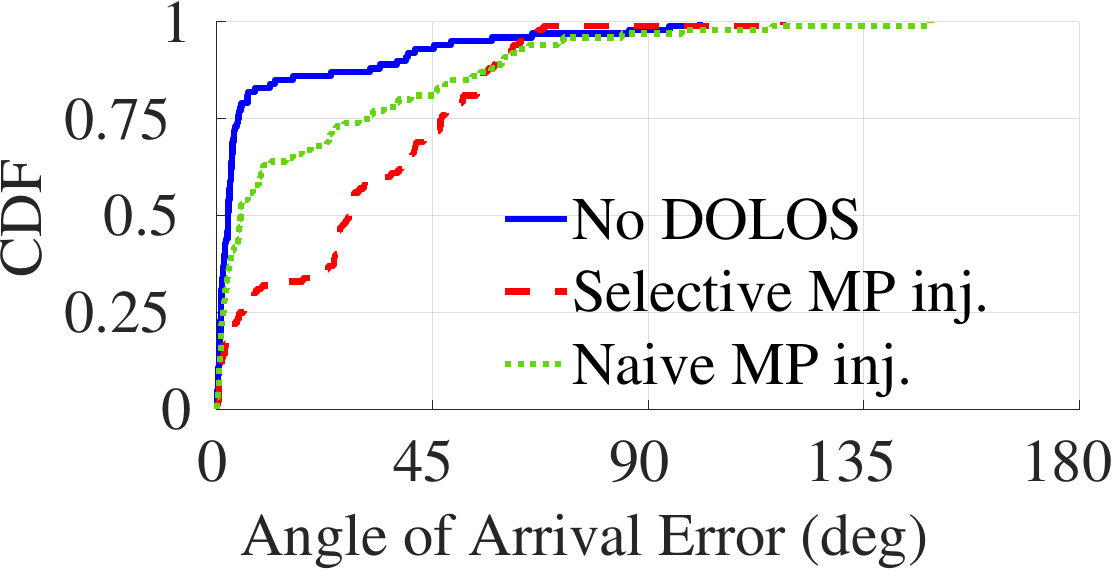}
        \caption{}
    \end{subfigure}
    \begin{subfigure}{0.30\linewidth}
        \includegraphics[width=\linewidth]{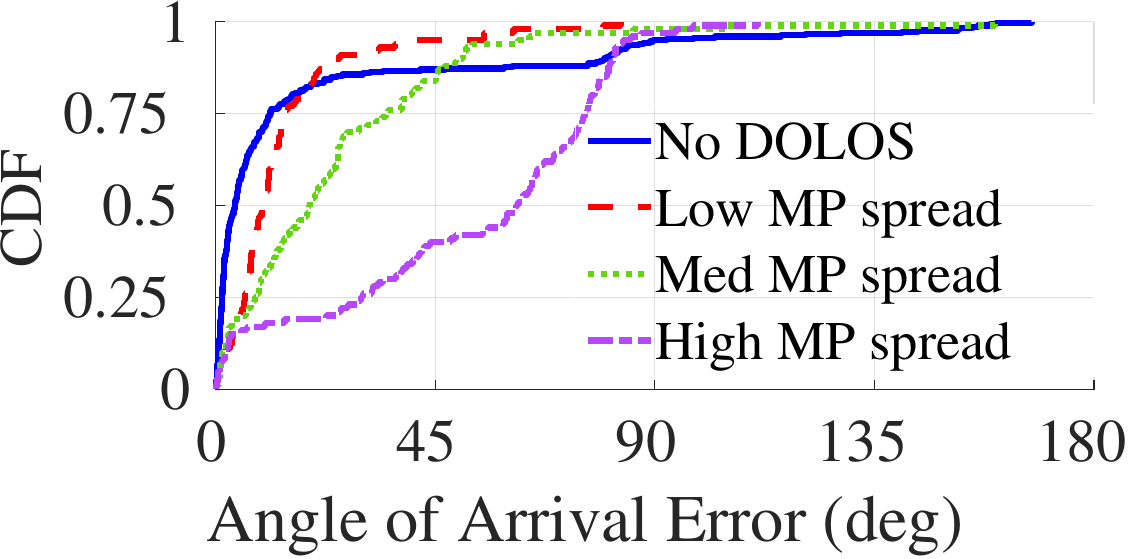}
        \caption{}
    \end{subfigure}
    \caption{(a) AoA estimation error when obfuscation delay is randomized. (b) AoA estimation error with and without selective multipath injection (c) AoA estimation error in a variety of multipath spreads.}
    \label{fig:mbs}
\end{figure*}

\subsubsection{Compensating direct path precoder} 

In \cref{sec:design-bf}, we discussed the need to apply additional compensation $\frac{1}{\alpha}\exp(-\iota \phi)$ to the direct path precoder. The optimum way to direct power towards the direct path and simultaneously null towards various reflections may not completely align in phase and magnitude to a purely transmitting in the direct path regime. The additional compensation provides this alignment.

In \cref{fig:comp-need}, we showcase the importance of this compensation parameter. Without applying any delay to the direct path, the direct path signal is shown at a distance of around 25m in \cref{fig:comp-need} (a). Without the explicit phase correction applied across the frequency subcarriers, we find multiple expressions of the direct path signal across the distance domain compared to the original channel in \cref{fig:comp-need}(b). This smearing of the direct path can undo the delays we seek to apply to the direct path. The additional compensation, however, removes this smearing effect as seen in \cref{fig:comp-need}(c). This is a consequence of the broadband nature of Wi-Fi signals. The direct path precoder we compute independently for each frequency introduces phase jumps across the subcarriers. As seen in \cref{sec:primer:localization}, a consistent phase increase across frequencies contributes to the ToF estimation of the direct path. Conversely, inconsistent phase jumps imply ambiguous ToF estimations, leading to the previously mentioned smearing effect.  

\subsubsection{Random delays to hide \name's obfuscation}

%As explored in \cref{sec:mirage-known}, a 
A simple workaround to discover the direct path AoA and ToF is to leverage the fact reflections in the environment change more rapidly than the direct path. An attacking AP can observe multiple packets transmitted by the user, average them, and then localize the user. Similar techniques are utilized in past works~\cite{arraytrack} to obtain better localization accuracy. \name must also continue to obfuscate the user's locations under these scenarios. To illustrate the need for random delays, we implemented the averaging within the simulator and quantified the results. 

In \cref{fig:mbs}(a), when \name applies constant delays to the direct path, we can see that the obfuscation in AoA estimation is reduced, i.e., an AP can more accurately predict the AoA of the received packet, indicating the degradation in \name's performance. However, \name can ensure inconsistent direct path representation across packets by applying random delays. This inconsistency improves obfuscation performance by $7.5\times$ when compared to applying contact delays, as seen in the red-dashed curve in \cref{fig:mbs}(a), and prevents an AP from accurately estimating the AoA.

\subsubsection{Selective multipath injection}

\cref{sec:add-refs} stressed the importance of reintroducing reflections through the precoder. This is needed to obfuscate the direct path amongst the reflections in the environment. However, we must also reintroduce the reflections without affecting the selective delays we apply to the direct path. Naively injecting reflections, or multipath (MP), directs power towards the direct path. This leaked power will arrive the earliest at the AP, allowing it to estimate the AoA accurately, and leading to loss of obfuscation. Instead, the obfuscation performance is preserved if we ``selectively'' reconstruct the multipath whilst nulling in the direct path regime. We simulate the transmission of hundreds of packets within simulation with varying multipath and selectively or naively inject multipath into the precoding vector. We then estimate the AoA for each of these cases as well as when no obfuscation is applied. \cref{fig:mbs}(b) showcases these results where ``Naive MP injection'' leads to degraded obfuscation performance and improves AoA estimation at AP by $5.5\times$ as compared to ``selective MP injection''. 

\subsubsection{Obfuscation in different multipath spreads}

\name's performance leverages the reflections in the environment to create ambiguity for the attacking AP. Consequently, the extent of environmental reflections can impact the obfuscation performance provided by \name. In the extreme case where there are no reflections in the environment, \name will fail to obfuscate the direct path AoA and ToF estimation. However, this is often not the case in indoor environments~\cite{soltanaghaei2018multipath}. A secondary impact on \name's performance is the extent of the environmental multipath spread. Put simply, if the reflections in the environment are arriving at the attacking AP in similar directions as the direct path, the errors in AoA obfuscation will be impacted to a lesser extent. This scenario is referred to as ``lower multipath spread''. Conversely, if the reflections are arriving from distinct directions, there is a ``high multipath spread''. Building physical settings that create varying multipath spreads is difficult, so we rely on simulations to illustrate our point. In \cref{fig:mbs}(c), we can see in situations with low spread, wherein the reflections can be as close as $10$ degrees to the direct path, the AoA estimation is improved significantly. Alternatively, under medium (at least $20$ degrees) and high (at least $40$ degrees) spreads, the AoA estimation errors degrade by $2\times$ and $~6\times$ as compared to lower MP spreads.

\begin{figure}
    \begin{subfigure}{0.39\linewidth}
        \includegraphics[width=\linewidth]{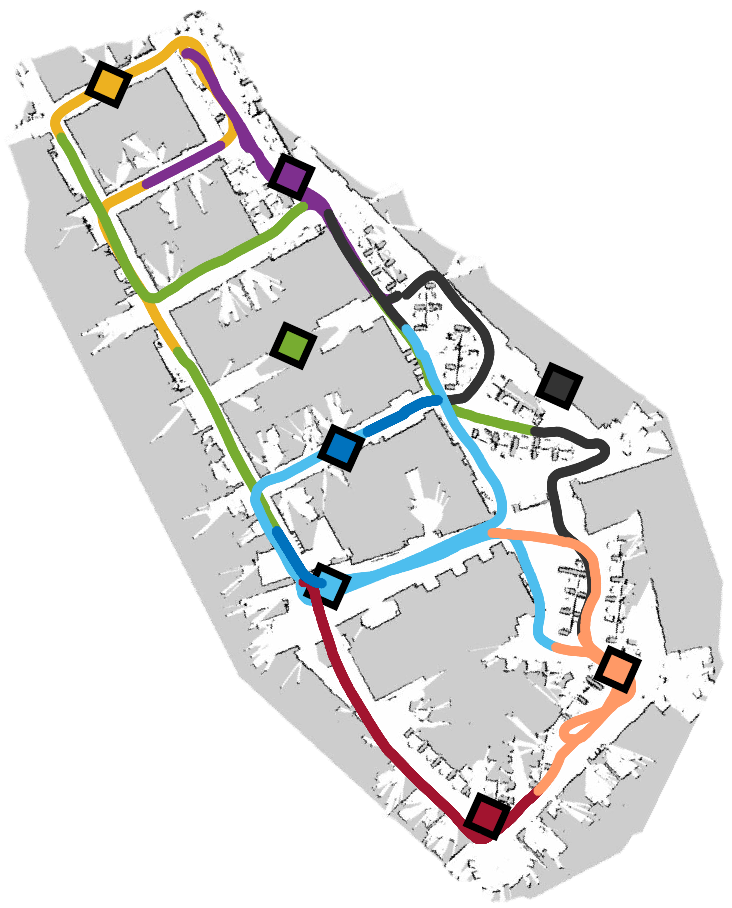} 
        \caption{}
    \end{subfigure}
    \begin{subfigure}{0.60\linewidth}
        \includegraphics[width=\linewidth]{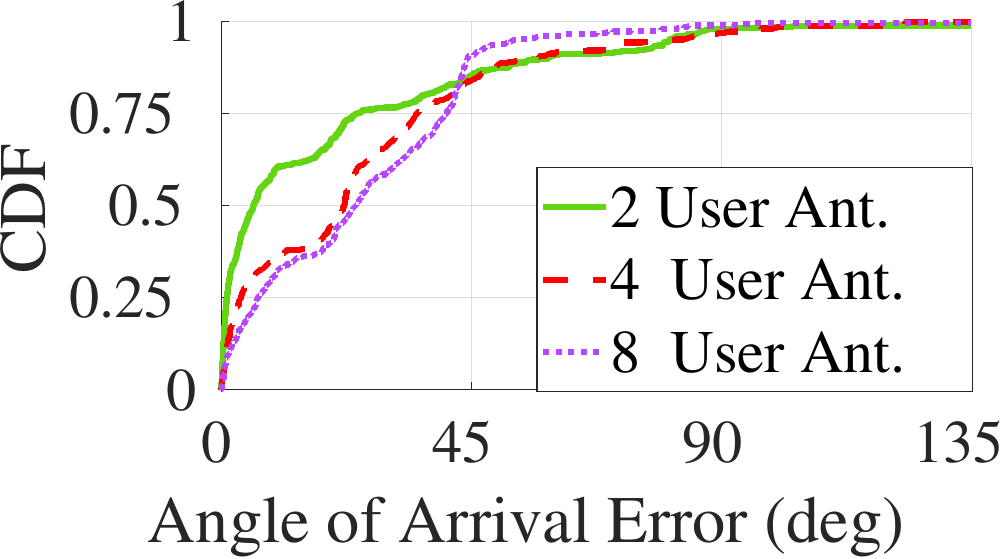}
        \caption{}
    \end{subfigure}
    \caption{(a) Wi-Fi user and AP association in an enterprise Wi-Fi network. The colored squares indicate the Wi-Fi APs deployed in an office building. The colored lines indicate that the Wi-Fi user automatically detects and connects to the nearest Wi-Fi AP. (b) AoA estimation obfuscation with varying number of user-side antennas}
    \label{fig:user:ap:association}
\end{figure}

\subsubsection{Wi-Fi user-AP association}\label{sec:eval-ap-assc}
In an enterprise network, the Wi-Fi APs are optimally deployed to provide efficient Wi-Fi user-AP association. As such, each Wi-Fi user will only be seen by one Wi-Fi AP. To demonstrate this, the Wi-Fi user walks around on the fifth floor of an office building as shown in \cref{fig:user:ap:association} (a). Eight Wi-Fi APs are deployed on this floor indicated as the colored squares. The colored lines in the figure indicate the walking path of the Wi-Fi user and which Wi-Fi AP he or she is associated with. As we can see, along his/her walking path, he/she can only connect with one Wi-Fi AP. This is because the deployment of Wi-Fi APs in the enterprise network is usually optimized to provide efficient connection. As such, \name can be efficiently used for location obfuscation in the enterprise Wi-Fi network.

\subsubsection{Impact of antenna array size at user device}
Since our location obfuscation algorithm mainly applies the precoding matrix to the Wi-Fi user for beamforming, its performance will be affected by the antenna array size. To demonstrate the impact of the antenna array size at the user device on the AoA estimation error after applying \name, we vary the antenna array size at the Wi-Fi user under a scenario where we assume medium multipath spreads (i.e., at least 20 degrees) in the simulation. We can see that the AoA estimation error increases as the antenna array size at the user increases shown in \cref{fig:user:ap:association}(b). Specifically, the median AoA estimation errors with 4 and 8-user antennas increase by 3.8$\times$ and 4.2$\times$ as compared to the 2-user antennas case. This is because more user antennas can provide higher beamforming resolution for location obfuscation. However, the median AoA estimation errors for 4 and 8-user antennas are similar. This is because the multipath diversity becomes dominant as the number of user antennas is beyond a certain threshold. Since the multipath diversity in the typical indoor wireless environment is usually limited within a certain threshold, the Wi-Fi user with 4 antennas designed in our \name is widely applicable and practical for location protection.

%{\color{red}ToDo: experiments on multiple-AP scenrios.}

%\subsubsection{RSSI Performance}
%It is important to maintain the wireless communication throughput while protecting the Wi-Fi user's location privacy. This requires our \name to not degrade the network throughput. 

%\noindent \textbf{Method.} To demonstrate this, we report the RSSI at the Wi-Fi AP, which can indicate the network performance of our system. We expect the RSSI estimated at the Wi-Fi AP should not be significantly degraded to maintain good wireless communication performance even after obfuscating the Wi-Fi user's location. 

%\noindent \textbf{Result.} Fig.~\ref{fig:end-to-end}(c) shows the CDF of RSSI without \name, with beamforming and delaying, with beamforming-nulling and delaying. As we can see, the RSSI is quite similar with or without obfuscation, which is around -76dBm. Even though \name obfuscates the direct-path signals, the wireless communication link is not significantly affected. This is because the wireless communication signals is going through the multipath environment, which can explore the multiplexing gain introduced by the space diversity. 

% !TEX root = paper.tex

\section{Related Work}\label{sec:related}
%{\color{blue} ToDo: "On the Security of Carrier Phase-based Ranging", https://arxiv.org/abs/1610.06077 "Security Assessment of Phase-Based Ranging Systems in a Multipath Environment", https://dl.acm.org/doi/pdf/10.1145/3517809 "Analog Physical-Layer Relay Attacks with Application to Bluetooth and Phase-Based Ranging", https://arxiv.org/abs/2202.06554} {\color{red} checked}

%The prior works on privacy-preserving Wi-Fi communication and sensing mainly introduce either smart surfaces or wireless radios to obfuscate the received wireless signals at the receiver, such that the receiver cannot analyze the wireless signals for sensing purposes. However, these works not only contradict the original purposes of wireless communication services for Wi-Fi users, often impacting the communication throughput but also require the extra hardware components that hinder ubiquitous deployment.

\noindent \textbf{Obfuscation through extra wireless radios:} Since the wireless sensing-based localization privacy threats~\cite{lu2024imperceptible} mainly focus on analyzing the wireless signals received at the receiver, it is very straightforward to introduce the extra wireless radios to either distort or relay the wireless signals such that location privacy is preserved. For example, Aegis~\cite{yao2018aegis} deploys an additional wireless radio that is instrumented with an amplifier and antennas to obfuscate the wireless signals. PhyCloak~\cite{qiao2016phycloak} proposes to introduce an extra full duplex radio to introduce the extra phase shift to the impinged wireless signals. %The extra phase shift introduced to the wireless signals will obfuscate the received signal's phase information at the attacker/receiver, such that the attacker cannot accurately localize the subject of interest. 
WiAdv~\cite{zhou2022wiadv} also leverages the extra wireless radio to introduce adversarial examples for machine learning-based wireless sensing systems.

\noindent \textbf{Obfuscation through smart surfaces:} %Active wireless radios can potentially interfere with other communication radios in high-density environments. Therefore, we can simply deploy the smart surfaces in the wireless environment to distort the wireless signals for privacy protection. 
The deployment of smart surfaces (or intelligent reflective surfaces, IRS) creates additional paths in the environment which can help corrupt the channel measured at the attacking AP. For example, authors in~\cite{zhu2018tu, sun2022sok} present geofencing with electromagnetic shielding paint on the walls or jamming to obfuscate the signal strength and further disable the signal strength-based indoor localization. These geofencing approaches require heavy workloads and cannot be widely deployed. %To further intelligently introduce the obfuscation to the wireless signals, 
RF-Protect~\cite{shenoy2022rf} introduces mmWave-based reflectors in the indoor environment to obfuscate the mmWave radar-based wireless sensing. However, extending this system to Wi-Fi frequencies and relying on their privacy mechanism will detrimentally impact the communication throughput as it introduces random phase offset to the signal, which may lead to destructive interferences. IRshield~\cite{staat2022irshield} instead proposes to use the smart surface in the Wi-Fi frequencies specifically to introduce phase noise into the system. However, multiple of these IRSs need to be deployed to provide widescale protection.

Recently, Fine Timing Measurement (FTM) protocol~\cite{ibrahim2018verification, yu2022precise,yu2020precise,jiokeng2020ftm} has been introduced to 802.11 standards for time-of-flight measurements that can be used for indoor localization~\cite{schepers2022privacy,ftm_secure,wipeep,schepers2021here}. For example, WiPeep~\cite{wipeep} exploits the loopholes in the standard 802.11 protocol to elicit the response from the existing Wi-Fi network and further harnesses the FTM protocol to achieve localization. To achieve location privacy protection, it is important to defend against these attacks by formally analyzing the Wi-Fi protocols. RAFA~\cite{liu2023exploring} exploits the privacy issues in machine learning-enabled wireless sensing systems with a focus on machine learning models, while our innovative precoding techniques can be applied to any wireless sensing system. Moreover, our \name design can fundamentally defend against these Wi-Fi protocols with privacy threats through AoA obfuscation and provides a software-only solution. The prior works discussed in~\cite{ma2019wifi} on exploiting the security issues of the carrier phase information in the wireless signals-based ranging, which is just the subcase of our precoding-based obfuscation approach.

\noindent \textbf{Location privacy protection at higher layers:} The location privacy can also be protected with higher-layer techniques~\cite{zhong2004privacy}. 
%Starting from the MAC-layer location privacy protection approaches,
For example, we can simply randomize the MAC address, or even provide a privacy guarantee for the MAC address~\cite{alaggan2018privacy}. %As a result, the user's identity recognized by the MAC address is masked, even though the user can be localized. 
However, the chip-level imperfections introduced by the radio devices can be used for fingerprinting and identifying the user, even though the MAC address is randomized. Recent paper~\cite{givehchian2023practical} has demonstrated that these hardware imperfections can be obfuscated to protect the user's location privacy, however, it only focuses on the BLE chipsets. It's not clear how it can be generalized to ubiquitous IoT devices. The application-layer approach~\cite{sung2020protecting} is also proposed to protect location privacy while the utility of the wireless communication service is compromised. However, \name is securing the location privacy from the physical layer perspective that can fundamentally protect the location privacy of the user. 

\looseness-1
% !TEX root = paper.tex
\section{Discussion, Limitations, and Future Work}\label{sec:conlusion}

In this paper, we show \name, a defense against the most common attack model for user device localization that uses commodity WiFi access points. We assume a setup where the commodity WiFi access points are compromised by an external attacker to get the user's device location without a user's consent. \name's software-only solution is akin to turning off GPS on your smartphone, to be used to obfuscate your locations while in public Wi-Fi networks. \name achieves this obfuscation by creating ambiguity about the user's locations at the attacking AP using the various reflections in the environment. With \name's precoding activated prior to transmission, this attack results in a $2.5\times$ increase in localization errors. 
%the attacking AP will assume the user's location based on the information of these reflected paths, subsequently resulting in a $2.5\times$ increase in localization errors. 
There are interesting challenges and opportunities that we look forward to:

\noindent\textbf{Monitoring Wi-Fi APs:} In dense AP deployments, multiple WiFi APs may monitor user's location. However, our current hardware is limited to obfuscating locations from monitoring APs due to a limited number of antennas. 
% Future devices instrumented with larger arrays can simultaneously extend the ideas presented here to obfuscate locations for multiple APs.   

% \noindent\textbf{RF based indoor Localization:} There has been great interest in UWB~\cite{zhao2021uloc} and BLE~\cite{botler2020direction} based indoor localization where there are slowly increasing deployments in industrial settings. \name can easily translate to these technologies -- for example, BLE has a whitening filter that we can modify further to create these obfuscations, similarly we can apply some precoding filters to UWB transmissions to filter out UWB-based AoA estimation modules~\cite{zhao2021uloc}. Further 802.11az~\cite{802.11az} \todo{complete here}

% \noindent\textbf{Trusted User Authentication:} Many important and authentic user applications like navigation, and tracking would need user location access, and we can see that using \name would obfuscate the user location at the attacking access point. We also want to ensure that an authorized access network can still locate the user. To that end, there are potential opportunities to look into encrypted authentication -- for example, to use the precoding matrix estimated in Eq~\ref{eq:prec-bf-delay} as the encryption key that could be sent over to the authorized access network, to decrypt the user location as needed.
\noindent\textbf{Trusted User Authentication:} There are potential opportunities to look into encrypted authentication -- for example, to use the precoding matrix estimated in Eq~\ref{eq:prec-bf-delay} as the encryption key that could be sent over to the authorized access network, to decrypt the user location as needed.

\bibliographystyle{abbrv}

\bibliography{reference}

%%%%%%%%%%%%%%%%%%%%%%%%%%%%%%%%%%%%%%%%%%%%%%%%%%%%%%%%%%%%%%%%%%%%%%%%%%%%%%%%
\end{document}